# Element Abundances and the Physics of Solar Energetic Particles

*Donald V. Reames*

Institute for Physical Science and Technology, University of Maryland, College Park, MD, USA

orcid.org/0000-0001-9048-822X

dvreames@gmail.com

**Abstract** Acceleration and transport of solar energetic particles (SEPs) causes their abundances, measured at constant velocity, to be enhanced or suppressed as a function of each ion's magnetic rigidity, and hence it's atomic mass-to-charge ratio $A/Q$. Ion charges, in turn, depend upon source electron temperature. In small "impulsive" SEP events, arising from solar jets, acceleration during magnetic reconnection causes steep power-law abundance enhancements. These impulsive SEP events can have 1000-fold enhancements of heavy elements from sources at ~2.5 MK, and similar enhancements of $^3$He/$^4$He and of streaming electrons that drive type-III radio bursts. Gamma-ray lines show that solar flares also accelerate $^3$He-rich ions but their electrons and ions remain trapped on magnetic loops so they dissipate their energy as X-rays, $\gamma$-rays, heat, and light. "Gradual" SEPs accelerated at shock waves, driven by fast coronal mass ejections (CMEs), can show power-law abundance enhancements or depressions, even with seed ions from the ambient solar corona. In addition, shocks can reaccelerate seed particles from residual impulsive SEPs with their pre-existing signature heavy-ion enhancements. Different patterns of abundance often show that heavy elements are dominated by a different source from that of H and He. Nevertheless, the SEP abundances averaged over many large events define the abundances of the corona itself, which is found to differ from the solar photosphere as a function of the first ionization potential (FIP) since ions, with FIP < 10 eV, are driven upward by forces of electromagnetic waves which neutral atoms, with FIP >10 eV, cannot feel. Thus, SEPs provide a measurement of element abundances in the solar corona, distinct from the solar wind, and may even better define the photosphere for some elements.

Keywords: solar energetic particles, solar system abundances, solar jets, solar flares, shock waves, coronal mass ejections, heliosphere.

## 1    Introduction

The relative abundance of chemical elements in any sample of material can be a clue to the identity and origin of that sample and to the nature of physical processes it has undergone. Energetic particles are no exception. Abundances reveal the age of the galactic cosmic rays (GCR) and the origin of unusual ions trapped in planetary magnetospheres. Solar energetic particles (SEPs) also display unique signature patterns of abundances that distinguish the physical processes that have formed them and a history they have traversed (Reames, 1988, 1999, 2013, 2021a, b, c).

A most unusual feature of SEP abundances is the small $^3$He-rich events with 1000-fold enhancements of $^3$He/$^4$He, later found to have enhancements of heavy elements extending as powers of the element atomic mass-to-charge ratio $A/Q$ from C and O up to elements as heavy as Pb, also by a factor of ~1000 (e.g. Reames et al. 2014a). These SEP ions have been associated with magnetic reconnection in solar jets and flares (Kahler et al. 2001; Reames, 2013, 2021a, b, c; Bučík, 2020). In some larger SEP events, these "impulsive-SEP" abundances often emerge as a signature of residual impulsive ions that have been reaccelerated by shock waves and exceed the average coronal abundances in some "gradual" SEP events (Desai and Giacalone 2016; Reames 1999, 2021a, b). Not only does this divide element abundances of SEP events into haves and have-nots, but it highlights variations in H and He which can either participate or not in the heavy-element behavior (Reames 2022b).

A most fundamental population SEPs can measure is the abundance of elements in the corona itself, from which all SEPs are derived. Only in the corona are densities low enough for ions to be accelerated without immediately losing their energy in Coulomb collisions. The corona provides a baseline for identifying other populations derived from it, but it also highlights the physical process that distinguishes it from the photosphere. We first discuss these reference coronal abundances, then the unique abundances of impulsive SEPs and then their presence or absence in the largest "gradual" SEP events. SEP abundances in this work are mainly derived from the low-energy matrix telescope (LEMT) on the *Wind* spacecraft (von Rosenvinge et al., 1995); ~30 years of LEMT abundance data are available from https://omniweb.gsfc.nasa.gov/ .

## 2    Reference Abundances, the Solar Corona, and FIP

The abundances of elements C, O, and above, in solar energetic particles (SEPs), were first measured using nuclear emulsion detectors on sounding rockets from Ft. Churchill, Manitoba by Fichtel and Guss (1961) and those measurements were later extended up through Fe using the same technique (Bertsch et al. 1969). As satellite measurements became available (e.g. Teegarden et al. 1973) comparisons of SEP with other abundances became more common (e.g. Webber 1975). Meyer (1985) summarized SEP abundance measurements in large SEP events as having a common baseline, derived from abundances in the solar corona where acceleration occurs, and a second component that varied from event to event as a power law in the particle charge-to-mass ratio $Q/A$ (Breneman and Stone 1985). A factor in this abundance variation was pitch-angle scattering of the ions; thus if Fe scatters less than O, Fe/O will be enhanced early in events, but depleted later. Such variations might average out, so it was soon possible to average ~50 or so large SEP events to remove the event-to-event variations and produce estimates of the coronal abundances (Reames 1995a, 2014); this could be compared with the solar photospheric abundances measured spectroscopically. A modern comparison of the SEP/photospheric abundances vs. first ionization potential (FIP) is shown in **Figure 1**. The photospheric abundances are the Caffau et al. (2011) modification of the Lodders et al. (2009) meteoritic abundances. A comparison using photospheric abundances of Asplund et al. (2021) is shown and abundances are listed by Reames (2021b).

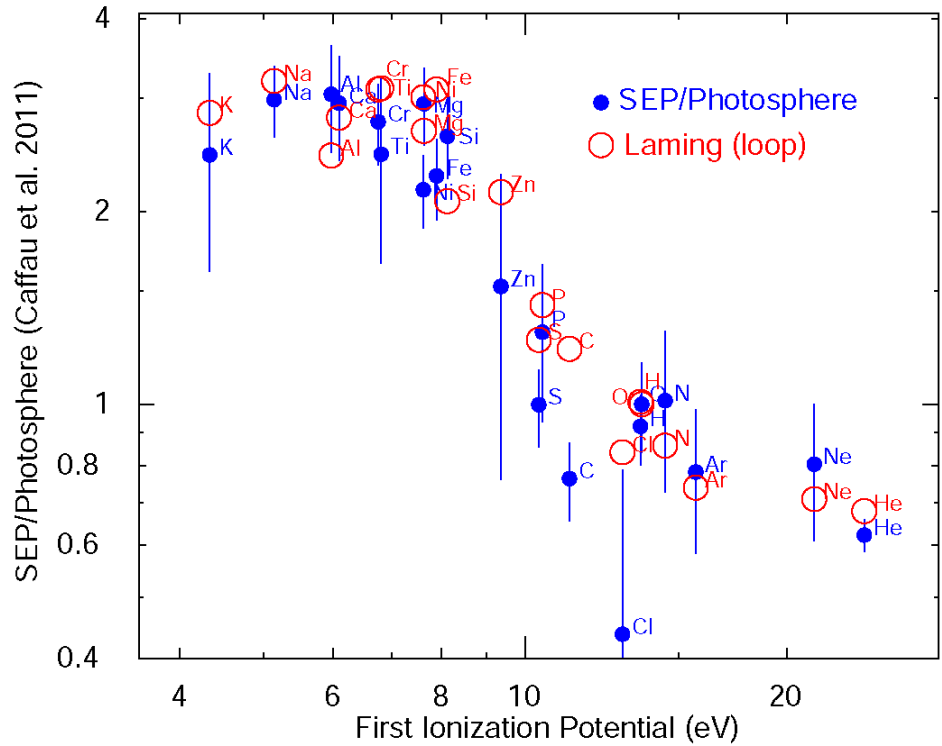

**FIGURE 1** The ratio of SEP to photospheric abundances of elements is shown vs. first ionization potential (**solid blue circles**) and is compared with theory of Laming et al. (2019; **open red circles**) for element transport along closed loops in active regions.

In the theory of Laming (2015; Laming et al. 2019) shown in **Figure 1**, the ponderomotive force of Alfvén waves helps drive low-FIP ions up across the chromosphere into the corona, but cannot affect un-ionized high-FIP neutral atoms. All elements become ionized in the hot ~1 MK corona. SEPs have a different FIP pattern than the solar wind or the solar wind accelerated by shock waves at co-rotating stream interfaces (Reames et al. 1991; Mewaldt et al., 2002; Reames 1995a; 2018a; 2021a); thus SEP abundances do not differ when measured in fast or slow wind (Kahler et al., 2009). SEPs are *not* accelerated solar wind. Differences between the FIP patterns of SEPs and the solar wind may be caused by open vs. closed field lines where Alfvén waves resonate with the loop length of closed loops (Reames 2018a; Laming et al. 2019). A first-order examination of **Figure 1** shows reasonable agreement when comparing elements with similar FIP but different $A/Q$, e.g. Mg or Si with Fe or Ni.

To what extent does this average of abundances over many gradual SEP events recapture coronal abundances? It is quite possible that the increasing and decreasing power-laws of abundances vs. $A/Q$ do not perfectly average out; however, a more outstanding disagreement seems to be the single element C. How can a single element, or actually a single ratio C/O, stand out? We will return to this question after discussing the patterns of known abundance variations and their probable causes.

## 3    Impulsive SEP events

The idea of two fundamental mechanisms of SEP acceleration began quite early (Wild et al. 1963) with solar radio observations that distinguished the sources of type II and type III radio bursts. Radio emission frequency depends upon the square root of the local electron density which decreases with distance from the Sun. Type III radio bursts have a rapid frequency decrease corresponding to the speed of 10 – 100 keV electrons streaming out from the Sun while type II bursts have the much slower frequency decrease of a ~1000 km s⁻¹ shock wave driven out from the Sun. These streaming electrons propagate scatter-free because the resonant turbulence that would scatter them is absorbed by the plasma (Tan et al. 2011). Later, Lin (1970) observed beams of ~40 keV electrons associated with the impulsive type III bursts and thought they might involve "pure" electron events, i.e. without ions. Relativistic electrons and energetic protons were only seen to accompany the shock-associated type II bursts.

Soon the SEP world was surprised by the observations of $^3$He-rich events. While a typical solar or solar-wind abundance is $^3$He/$^4$He ≈ 5 × 10$^{-4}$, an event was soon found with $^3$He/$^4$He = 1.5 ± 0.1 (Serlemitsos and Balasubrahmanyan, 1975; Mason, 2007). Such a high ratio could not come from fragmentation of $^4$He as occurred in GCRs since the $^3$He was not accompanied by any $^2$H. Later measurements of Be/O and B/O < 2 × 10$^{-4}$ (McGuire et al., 1979; Cook et al., 1984) completely laid to rest the early idea of nuclear fragmentation. Instead, this was a new mechanism involving resonant wave-particle interactions. The $^3$He gyrofrequency, dependent upon $Q/A$, lay isolated at $Q/A$ =2/3, between those of H at $Q/A$ =1 and $^4$He at $Q/A$ =1/2.

These two different features of impulsive SEPs, (1) "pure" electron beams producing type III radio bursts and (2) $^3$He-rich events, were unified by Reames et al. (1985) as different properties of the same events, and Reames and Stone (1986) explored kilometric radio properties of $^3$He-rich events, even tracking the flow of electrons out from the Sun. Early theories discussed selective heating by resonant absorption of various types of plasma waves at the $^3$He gyrofrequency followed by acceleration of the thermal tails by some unspecified mechanism (e.g. Fisk 1978; see other references in Reames 2021c or 2023c) but Temerin and Roth (1993) proposed electromagnetic ion cyclotron (EMIC) waves generated by the streaming electron beams and added their absorption by mirroring $^3$He for acceleration, in analogy with the production of ion conics seen in the Earth's magnetosphere.

## 3.1 Element Abundances

Enhancements of heavy elements up to Fe in impulsive events were first reported by Mogro-Comparo and Simpson (1972). These observations were improved in subsequent generations of experiments by Mason et al. (1986), then by Reames et al. (1994). Groups of elements were eventually resolved up to Pb at 3 – 10 MeV amu$^{-1}$ (Reames, 2000; Reames and Ng 2004; Reames et al., 2014) and below ~1 MeV amu$^{-1}$ (Mason et al., 2004). The average enhancement was found to be a power law in $A/Q$ with a power of 3.64 ±0.15 above 1 MeV amu$^{-1}$ and ~3.26 below, using $Q$ values appropriate for ~3 MK.

Early direct measurements of ionization states of SEP elements up to Fe (Luhn et al., 1984, 1987) found $Q_{Fe}$ =14.1 ± 0.2 in gradual SEP events, which would correspond to a typical source plasma temperature of ~2 MK, but $^3$He-rich events had $Q_{Fe}$ =20.5 ± 1.2 with elements up to Si fully ionized, suggesting either a temperature >10 MK, or the stripping of the ions by passing through a small amount of material after acceleration. Reames et al. (1994) noted that ion enhancements, relative to the corona, formed three groups (1) C, N, and O like $^4$He, all seemed to be un-enhanced, (2) Ne, Mg, and Si were enhanced about a similar factor of ~2.5, and (3) Fe was enhanced a factor of ~7. The first group was probably fully ionized with $A/Q = 2$ while the second group would have similar abundances in states with 2 orbital electrons, which occurs at about 3 MK. No enhancements could occur if Ne, Mg, and Si, were fully ionized, as measured. The resolution of this dilemma is that the ions in impulsive SEP events are stripped after acceleration, and it was later found that the ionization states of Fe depended upon ion velocity (DiFabio et al., 2008) as expected from stripping. This suggested that impulsive SEP events were accelerated at ~1.5 $R_S$. In contrast, gradual SEP events, found to be accelerated at shock waves driven out from the Sun by coronal mass ejections (CMEs), began at 2 – 3 $R_S$ (Tylka et al. 2003; Cliver et al., 2004; Reames 2009a, b).

### 3.2    Jets and Flares

While gradual SEP events were found to have a 96% correlation with fast, wide CMEs, (Kahler et al., 1984), an early search found no meaningful association of ³He-rich events with CMEs observed by the Solwind coronagraph (Kahler et al. 1985). However, with improved coronagraph sensitivity of SOHO/LASCO, Kahler et al. (2001) found narrow CMEs that were associated with the larger ³He-rich events; the CME associated with the large impulsive SEP event of 1 May 2000 had a speed of 1360 km s⁻¹, easily fast enough to drive a shock that would reaccelerate particles. These observations led Kahler et al. (2001) to associate impulsive SEP events with solar jets (e.g. Shimojo and Shibata 2000), an association that has been extended (Nitta et al. 2006; Wang et al., 2006; Bučík et al., 2018a, b) and reviewed by Bučík (2020). Jets are driven by energy from magnetic reconnection as shown in the sketch in **Figure 2**.

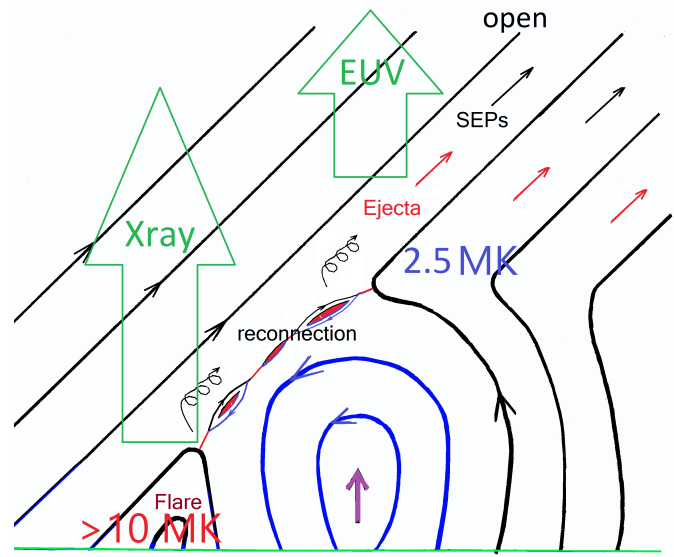

**FIGURE 2.** Sketches the topology of a solar jet where rising closed field lines of one magnetic polarity (**blue**) form islands of reconnection where they meet the oppositely-directed open field lines (**black**). SEPs are accelerated in the reconnection region and escape along the open field lines, as does CME plasma. Newly formed closed field lines at the lower left trap energetic electrons and ions that plunge into the deeper corona and form a flare. The hot flaring region (>10 MK) emits X-rays while the open region is observed to be an EUV-emitting region at ~2.5 MK (Reames 2023c).

A particle-in-cell simulation of a reconnection region by Drake et al. (2009) found strong $A/Q$-dependent enhancements in the energetic heavy ions that could match those observed. Particles were Fermi-accelerated as they reflected (mirrored) back and forth by the approaching ends of the collapsing islands of magnetic reconnection. However, the reconnection that opens some field lines always closes others as shown in the lower left of **Figure 2**. These closing field lines trap newly-accelerated particles that deposit their energy as heat or in emission of X-rays or γ-rays – a solar flare. Thus jets would always be expected to have accompanying flares that involve the same accelerated particles. Of course there are more realistic models of jets (e.g. Archontis and Hood, 2013; Lee et al., 2015; Pariat et al., 2015) that better describe the CME emission, but jet models do not yet include SEP acceleration.

The relationship of jets vs. flares is a close one, with similar ion acceleration on open vs. closed field lines. Similar ion enhancements were first noted between impulsive SEPs and abundances from γ-ray line measurements in large flares by Murphy et al. (1991), then Mandzhavidze et al. (1999) found that energetic ions accelerated and trapped in all 20 available large solar flares were ³He-rich. Unusually strong were the three γ-ray lines at 0.937, 1.04, and 1.08 MeV from the de-excitation of ¹⁹F* produced with especially high cross section in the reaction ¹⁶O (³He, p) ¹⁹F*. These were compared with other lines from excited ¹⁶O, ²⁰Ne, and ⁵⁶Fe, to distinguish ³He from ⁴He in the "beam." Some of the events had ³He/⁴He ~ 1 while all had ³He/⁴He > 0.1. Murphy et al. (2016) later found six key ratios of γ-ray fluxes dependent upon ³He/⁴He in the beam and all showed an average ³He/⁴He ratios of 0.05 – 3.0. These studies included ~135 product de-excitation lines from ~300

proton- and He-ion induced reactions (e.g. Kozlovsky et al. 2002). These $\gamma$-ray lines are from the largest flares, not small jet-associated impulsive flares, suggesting that the impulsive-SEP abundances are a general consequence of the physics of magnetic reconnection.

### 3.3 Power-Law Abundances from Jets With or Without Shocks

We always compare element abundances at the same velocity, or MeV amu$^{-1}$, but properties such as magnetic deflection and scattering depend upon magnetic rigidity, or momentum per unit charge, also depend upon $A/Q$, quite often upon a power of $A/Q$ (e.g. Parker 1965). Thus it is not surprising that enhancements that depart from the reference abundances vary as a power of $A/Q$. More specifically for impulsive SEPs, the theory of Drake et al. (2009) relates the power of $A/Q$ to the power of the width distribution of the reconnecting magnetic islands. However, the $Q$ values of the ions depend upon the source plasma temperature. Our strategy has been to simply try all temperatures in a reasonable range, determine the $Q$ values (using e.g. Mazzotta et al., 1998 or Post et al., 1977) and choose the temperature and power law that gives the best least-squares fit of enhancement vs. $A/Q$ (Reames et al., 2014b; Reames, 2018b, 2021a). Examples showing typical temperature dependence in enhancement vs. $A/Q$ have been illustrated in Figure 6 of Reames (2022b), Figure 6 of Reames (2018), and Figure 2 of Reames et al. (2014b).

Fitting 111 impulsive SEP events Reames et al. (2014b) found 79 events at 2.5 MK and 29 at the neighboring 3.2 MK, i.e. very little variation for impulsive SEP events. Subsequently these temperatures have been shown to agree with EUV temperatures in jets (Bučík et al., 2021). **Figure 3** shows power-law fits to abundance enhancements in several impulsive SEP events. Time profiles of the events are shown in the lower panels, derived temperatures (and event durations) in the middle panels and best fits to the enhancements vs. $A/Q$ in the upper panels. Event numbers marking each event onset in **Figure 3** correspond to the impulsive SEP event list in Reames et al. (2014a), all events selected to have enhanced Fe/O abundances.

For Events 3 and 4 in **Figure 3**, the power-law fits, obtained for ions with $Z \geq 6$, extend to include proton measurements at $A/Q = 1$. This is taken to mean that all of these ions come from the same population, i.e. the magnetic reconnection in the associated jet. Reames (2020a) defined these "pure" reconnection events as SEP1 events; they had either no visible CME or CME speeds < 500 km s$^{-1}$, i.e. no shock acceleration was likely. Event 5 is ambiguous; the excess protons do not fit the power law, but the theory of Drake et al. (2009) allows for enhancements that start above $Z = 2$, and H and $^4$He could both be un-enhanced, i.e. at the same level, as in this event, which also shows no CME.

There was a simpler time when there were thought to be only two types of SEP events, $^3$He-rich or "impulsive" SEP events with unique element abundances and shock-accelerated "gradual" SEP events that accelerated coronal ions. Abundances in the gradual events varied primarily because of differences in elements transport - since Fe scatters less than O, Fe/O will be enhanced early and depressed later, following a power-law in $A/Q$. This was observed by Breneman and Stone (1985) and implied in the discussions of Meyer (1985). This simplicity ended when Mason et al. (1999) found enhances $^3$He in a large gradual SEP event. Clearly, shocks could reaccelerate residual ions from small impulsive as well as ambient coronal ions and these two seed populations became widely discussed (Tylka et al. 2001, 2005; Desai et al., 2003; Tylka and Lee 2006). Eventually, Reames (2020a) suggested organizing the combination of acceleration mechanisms and seed populations into four physical categories shown in Table 1.

**Table 1** Properties of Four Sources of SEPs

|  | **Observed Properties** | **Physical Association** |
|---|---|---|
| **SEP1** | Fe-rich power-law enhancement vs. $A/Q$ at all $Z$ ; $T \approx 2.5$ MK | Magnetic reconnection in solar jets with no fast shock |
| **SEP2** | Fe-rich power-law enhancement vs. $A/Q$ at $Z > 2$; $T \approx 2.5$ MK. Proton excess ~×10. CME speed >500 km/s | Jets with fast, narrow CMEs drive shocks that reaccelerate local SEP1 seeds to dominate high $Z$ and ambient plasma to dominate H (and $^4$He) |
| **SEP3** | Fe-rich power-law enhancement vs. $A/Q$ at $Z > 2$; $T \approx 2.5$ MK. Proton excess ~×10. CME speed >> 500 km/s | Fast, wide CME-driven shocks accelerate SEP1 residue left by many jets in active-region pools, plus H (and $^4$He) from ambient plasma at low $Z$. |
| **SEP4** | Power-law or flat vs. $A/Q$ for all ions with 0.8< $T$ < 1.8 MK. Fast, wide CME | Very fast, wide CME-driven shocks accelerate all dominant ions as seeds from the ambient plasma. |

However, Event 92 in **Figure 3F** is an event with a large proton excess, and even an excess in $^4$He, suggesting power-law contributions that are shock-accelerated from two seed populations, ambient-coronal ions for H and $^4$He, and residual impulsive suprathermal ions for $Z \geq 6$. The event would be classified SEP2 if all the SEP1 ions came from a single impulsive jet event and SEP3 if the pre-accelerated impulsive seed ions had collected from many previous SEP1 or SEP2 events before shock acceleration. In fact this seems to be a SEP3 event, despite its short duration, since its intensity is high and an earlier event in the same location is similar in character. Note also that the scatter of the $Z \geq 6$ points about the fit line in **Figure 3F** is quite small compared with those in **Figure 3C**, suggesting that output of many small jets have been averaged to reduce the abundance variations in SEP3 events (see Figure 8 in Reames, 2020a). We will consider other SEP2 and SEP3 examples below.

The abundance of $^4$He in SEP events can vary because it can be dominated by either coronal or impulsive seed components. However, there are also other variations in $^4$He that have been summarized in greater detail by Reames (2022b).

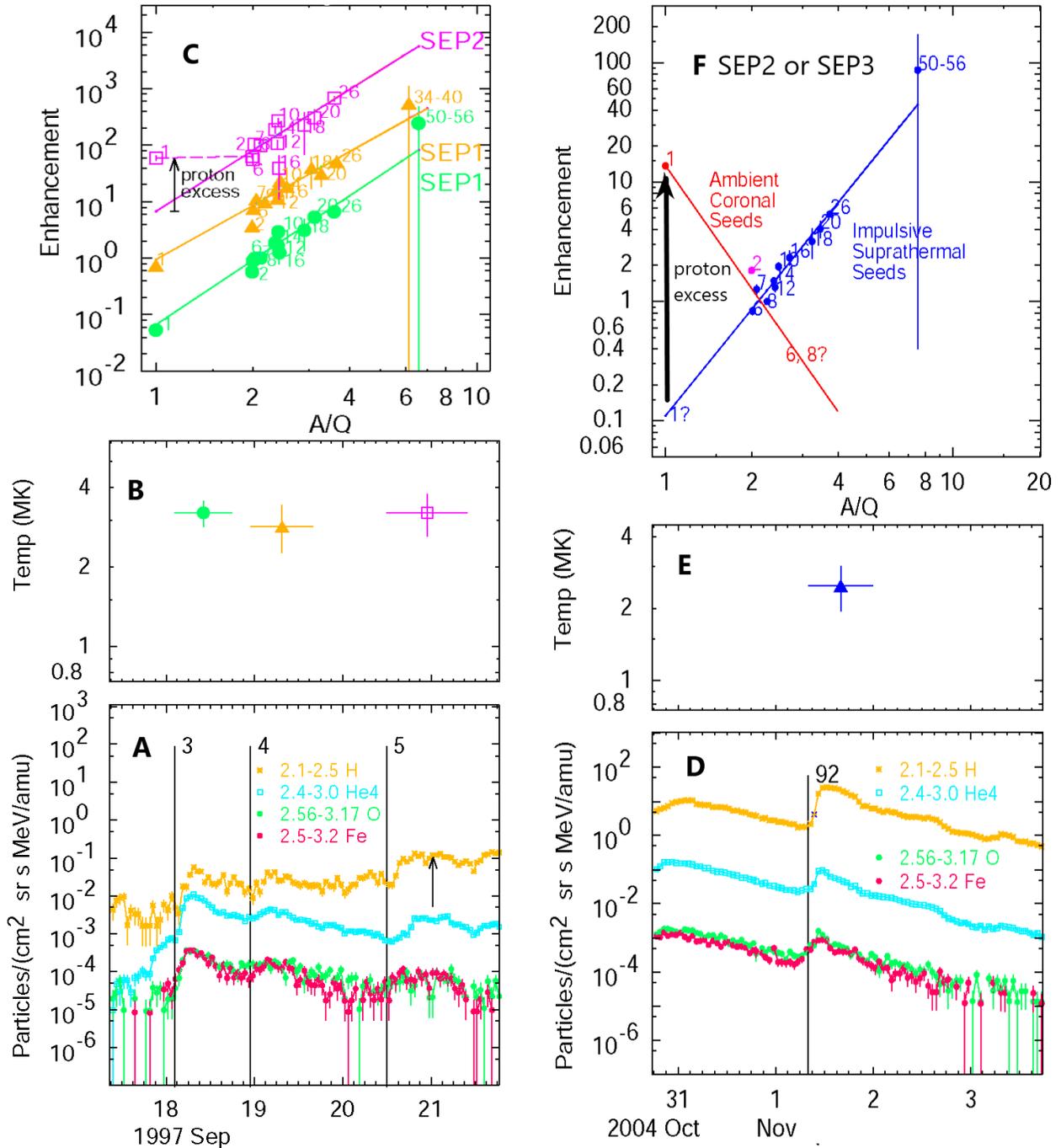

FIGURE 3 Panels **A** and **D** show intensities of H, $^4$He, O and Fe at the listed energies in MeV amu$^{-1}$ for **A** a sequence of three small impulsive SEP events and **D** a larger event. Panels **B** and **E** shows the derived best-fit temperatures (and durations) for each event below; panels **C** and **F** show the corresponding best-fit power-law abundance enhancements (normalized at O and shifted ×10 for each event), with the measurements for each element labeled by atomic numbers $Z$. Only elements with $Z \geq 6$ are included in the fits. Panel **F** distinguished two possible seed populations contributing to acceleration in the shock wave driven by the 925 km s$^{-1}$ associated CME (see text).

## 4    Gradual SEP events

The first recognized SEP events (Forbush 1946) were the immense ground-level enhancement (GLE) events where GeV protons initiate nuclear cascades through the atmosphere that exceed those of galactic cosmic rays (GCRs). While these events were erroneously attributed to solar flares for many years (Gosling, 1993, 1994), Kahler et al. (1984) had found that large SEP events had a 96% association with fast, wide shock waves, driven out from the Sun by coronal mass ejections (CMEs), reaffirming the finding of shock acceleration in radio type II bursts by Wild et al. (1963) two decades earlier. Mason et al. (1984) concluded that only "large-scale shock acceleration" could explain the extensive rigidity-independent spread of SEPs, and recent findings from missions like STEREO now show how shock waves and SEPs wrap around the Sun (e.g. Reames 2023a, b). Evidence for shock acceleration has grown (e.g. Reames 1995b, 1999, 2013, 2021b; Zank et al. 2000, 2007; Lee et al. 2012; Desai and Giacalone 2016; Kouloumvakos et al. 2019), especially for GLEs (Tylka and Dietrich 2009; Gopalswamy et al. 2012, 2013a; Mewaldt et al. 2012; Raukunen et al. 2018).

One line of evidence has been the onset timing or solar particle release (SPR) time of the SEPs compared with X-ray or $\gamma$-ray onset times of associated flares. Ions with lower velocities have increasingly delayed onsets that extrapolate back to a single SPR time, with the delay = distance along the observer's field line over velocity. For impulsive events the SPR times and X-ray onsets agree closely (Tylka et al. 2003) but for gradual events like GLEs the SPR time can lag the X-ray and $\gamma$-ray onset by as much as half an hour (Tylka et al. 2003; Reames 2009a, b), sometimes after the associated flare is completely over. The SPR time corresponds to the time the shock at the leading edge of the CME reaches $2 - 3$ solar radii (Reames 2009a, b; Cliver et al. 2004), presumably when the shock emerges above closed magnetic loops and its speed exceeds the declining Alfvén speed. Type II radio emission shows that shocks can begin at ~1.5 AU (Gopalswamy et al. 2013b) but the SPR time depends upon the observer's longitude which often differs from that of the closest source of radio emission (Reames 2009b). Variations in these parameters or delays in the shock interception of the observer's field line (Reames 2023a, b) cause variations in the SPR delay.

The abundance enhancements or suppressions in gradual SEP events relative to the reference coronal abundances have been classed as SEP3 or SEP4 by Reames (2020a) depending upon whether the source of seed particles for the shock acceleration is purely ambient coronal abundances (SEP4) or whether the residual impulsive suprathermal ions dominate the heavy elements (SEP3).

### 4.1 Moderate-sized SEP4 Events

**Figure 4** shows abundance data for two typical SEP4 events. These gradual events are intense enough to measure significant enhancements of most elements in several time intervals, e.g. every 8 hours. The derived temperatures are less than those in impulsive events and the abundance patterns vary little during the events, suggesting that there is too little variation in scattering to separate different elements in time. Most important, the power-law fits, obtained for the elements C and above can be extended to fit H and He reasonably well, suggesting that all the elements have come from a single population. This, the declining enhancement, and the lower temperature, suggests that that population is ambient coronal ions, un-enhanced by any impulsive pre-accelerated population. Similar plots can be seen in some events from three spacecraft (e.g. *Wind* and STEREO A and B) spaced at ~120º around the Sun (e.g. see Figure 7 in Reames, 2022b). The main reason for the systematic decline with $A/Q$ may be that higher-rigidity elements leak away faster. The energy spectra of ions are correlated with variation in $A/Q$ in these events (Reames 2021d, 2022a), and are also relatively steep.

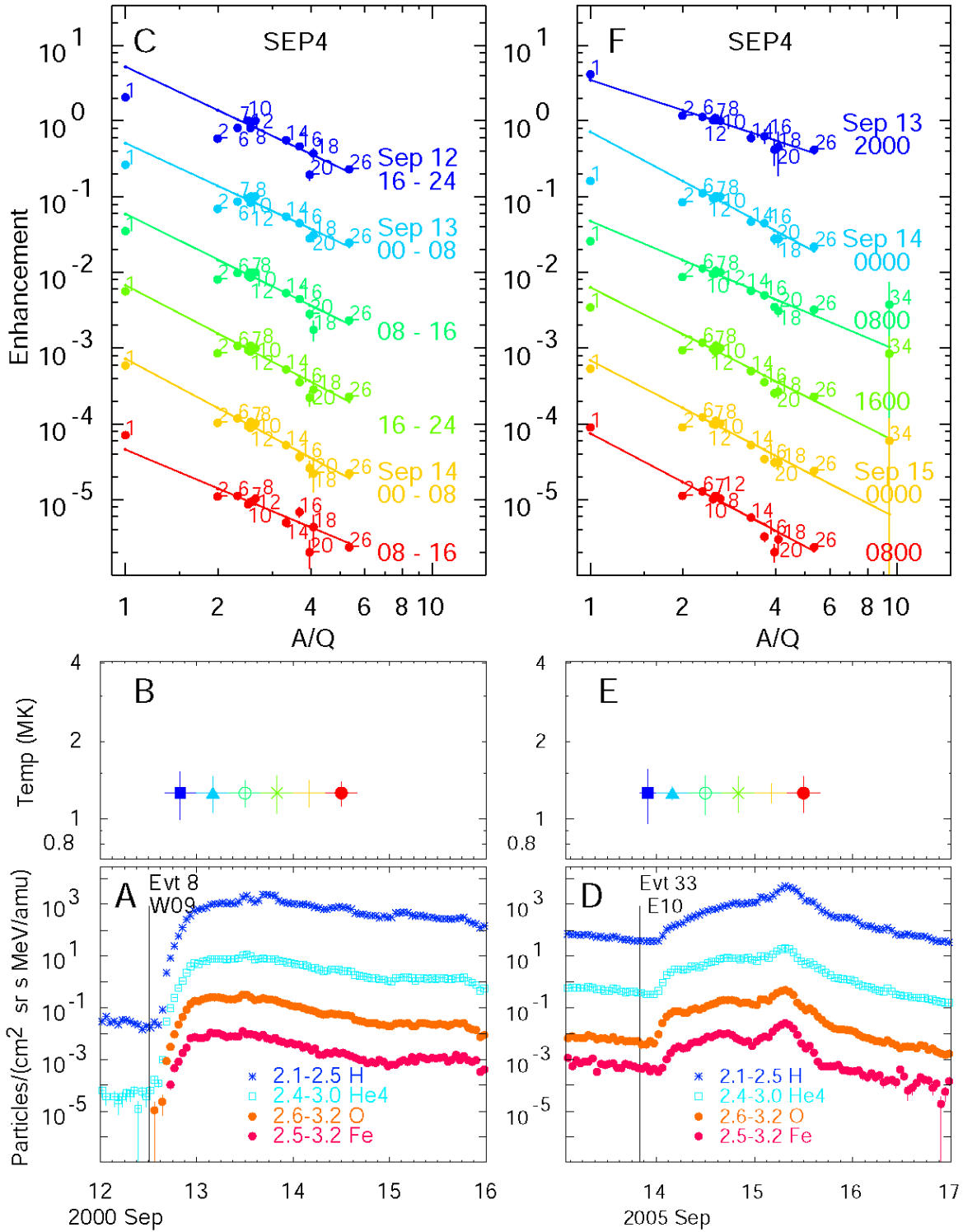

**FIGURE 4** Typical small gradual SEP4 events. Panels **A** and **D** show time evolution of the elements H, ⁴He, O and Fe and listed energies (in MeV amu⁻¹) in two gradual SEP events numbered 8 and 33 (in reference to the list in Reames 2016). Panels **B** and **E** show derived source plasma temperatures in a series of intervals (**colors**) during each event. Panels **C** and **F** show abundance enhancements vs. A/Q for elements (normalized at O and shifted ×0.1 for each interval) with atomic numbers Z shown for each time interval (**color**) and the best fits of elements with Z ≥ 6 extended to A/Q = 1. All the elements, including H, tend to fit each power law.

Typically, enhancements decline with $A/Q$ in many gradual events, as in **Figure 4,** but there is also a class of events where the $A/Q$ dependence is flat, i.e. the abundances are nearly coronal. However, they cannot be used to determine a temperature well, since the enhancements are independent of $A/Q$. Thus, these events are unremarkable and often overlooked. However, one has been included in **Figure 5,** as an example of an event where abundance enhancements (**Figure 5c**) begin as quite flat, i.e. independent of $A/Q$, but then steepen as the higher-rigidity ions preferentially leak away. Here, source temperatures are poorly determined when enhancements are flat, but improve as they steepen.

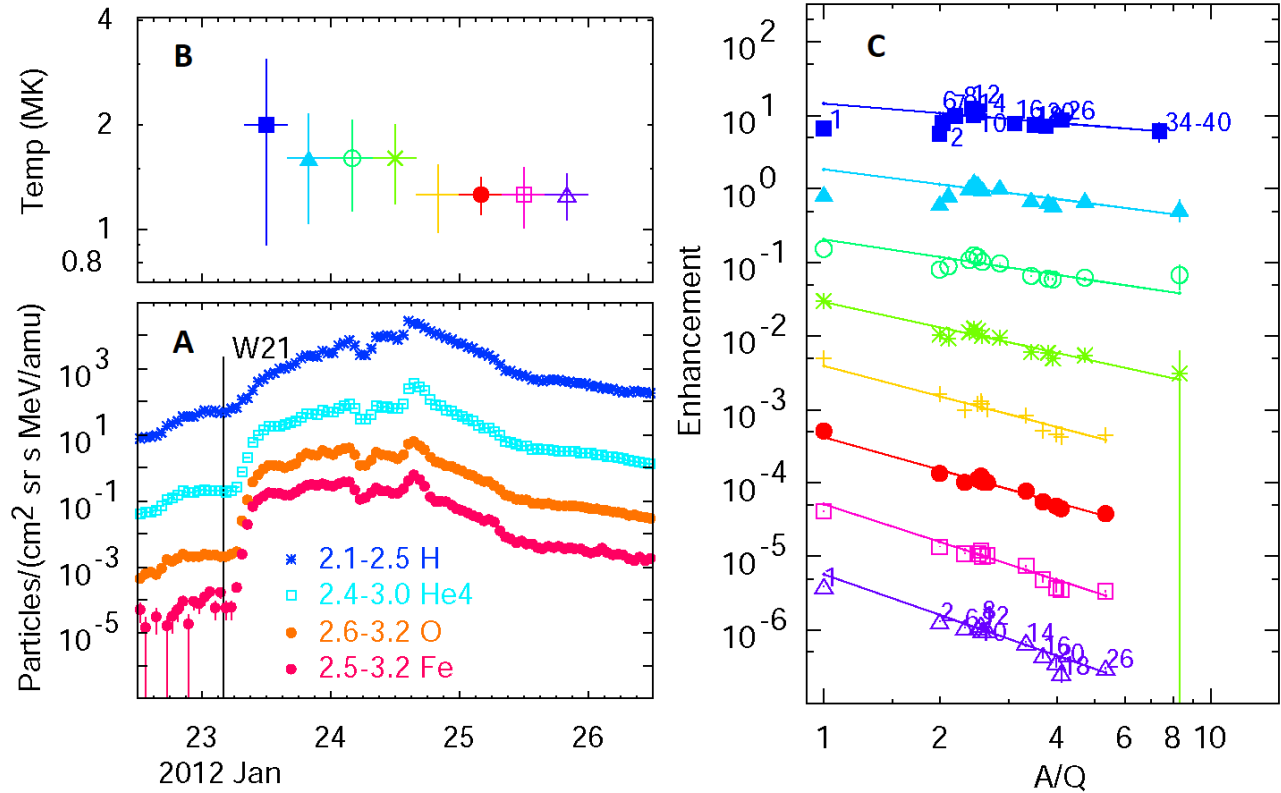

**FIGURE 5** A typical gradual SEP4 event with initially flat (coronal) abundances. Panel **A** shows time evolution of the elements H, ⁴He, O and Fe and listed energies (in MeV amu⁻¹) in the gradual SEP. Panel **B** shows derived source plasma temperatures in a series of intervals (**colors** and **symbols**) during the event. Panel **C** shows abundance enhancements vs. $A/Q$ for elements in each time interval (**color** and **symbols**) ×0.1 and the best fits of elements with $Z \geq 6$ extended to $A/Q = 1$. Element atomic numbers $Z$ are listed for the first and last interval.

### 4.2  GLEs that are SEP4 Events

As intensities in gradual SEP events increase, the high-energy protons stream ahead to amplify Alfvén waves that scatter subsequent ions (Stix, 1992; Ng et al. 1999, 2003, 2012). The scattering varies as a power law in magnetic rigidity so that Fe can propagate away from the shock more easily than O, for example. **Figure 6** shows abundance fits for two GLEs that are SEP4 events. The fitted temperatures are near ~1 MK and the fits for elements with $Z \geq 6$ are in reasonable agreement with H and He for both positive and negative power-law slopes. Heavy elements tend to be enhanced ahead of the shock and hence depressed behind. These enhancements come primarily from strengthened preferential scattering during transport in these events, *not* from impulsive seed particles, as indicated by the lower temperature, the inclusion of protons in a single seed population, and the return to heavy-element suppression behind the shock. Where the abundances show an increase with $A/Q$ the energy spectra have become quite flat in the plateau region (Reames and Ng 2010, 2014; Ng et al. 2012; Reames 2021a) because of the underlying correlation between abundances and spectra (Reames 2021d, 2022a).

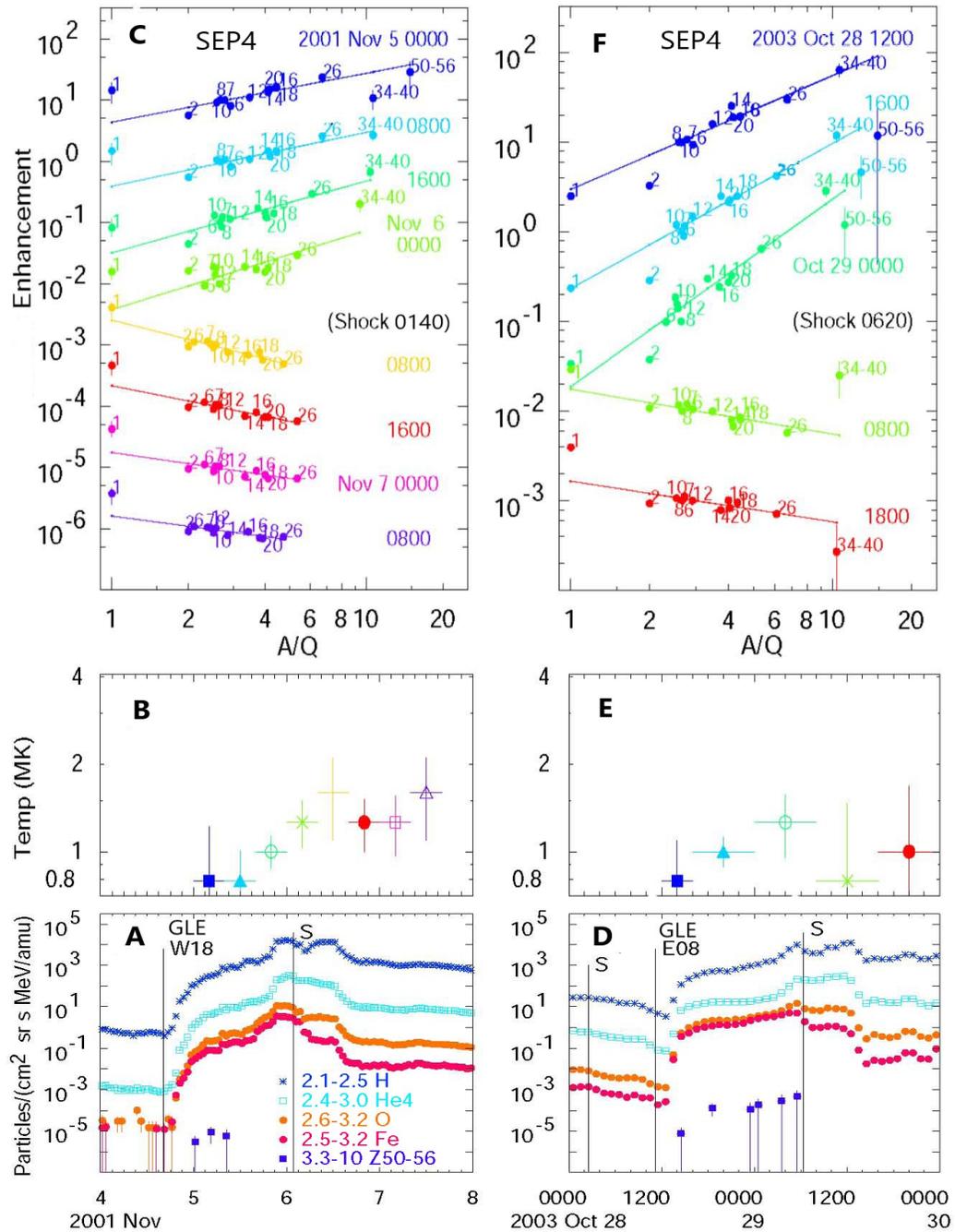

**FIGURE 6** Large GLEs as gradual SEP4 events. Panels **A** and **D** show time evolution of the elements H, ⁴He, O, Fe, and elements with $50 \leq Z \leq 56$, along at the listed energies (in MeV amu⁻¹) in two SEP4 events. Panels **B** and **E** show derived source-plasma temperatures in a series of intervals (**colors**) during each event. Panels **C** and **F** show abundance enhancements vs. $A/Q$ for elements (normalized to O and shifted ×0.1 for each interval) with $Z$ shown, for each time interval (**color**) and the best least-squares fits of elements with $Z \geq 6$ extended to $A/Q = 1$. The high-$Z$ enhancements go away after the shocks pass.

For these large events transport out to 1 AU becomes much more important. The strong transport-induced abundance increases in **Figure 6** are a consequence of the high SEP intensities. High intensities of protons streaming away from the shock amplify resonant Alfvén waves (Stix 1992; Melrose 1980) that scatter subsequent ions. The wave number of resonant waves, $k \approx B/\mu P$ where $B$ is the magnetic field intensity, $P$ is the particle rigidity or momentum per unit charge, and $\mu$ is the cosine of the particle's pitch angle relative to $B$. The spectrum of waves not only traps particles near the shock but extends far out into space, bounding intensities at the "streaming limit" (Reames and Ng 1998, 2010, 2014), thus driving more acceleration (Lee 1983, 2005; Zank et al. 2000; Ng and Reames 2008; Afanasiev et al. 2015, 2023) and the transport strongly favors the escape of Fe vs. O (Parker 1965; Ng et al., 1999, 2003, 2012; Tylka et al., 2001, 2005; Tylka and Lee 2006). In very large events intensities of ions below ~1 MeV amu$^{-1}$ (and their abundances) remain at pre-event background until the shock comes very near the observer.

### 4.3    GLEs that are SEP3 Events

However, large gradual SEP events, even GLEs, can pick up pre-accelerated residual impulsive seed particles from multi-jet collections often observed to accumulate near active regions (Desai et al., 2003; Wiedenbeck et al., 2008; Bučík et al., 2014, 2015; Chen et al., 2015; Reames 2022a). Recently, Kouloumvakos et al. (2023) have found an average connection time to $^3$He-rich active regions of 4.1 ± 1.8 days, suggesting a width of ~52º in longitude. These seed particles contribute with their characteristic enhancement pattern at high $Z$ and its source temperature, but ambient coronal ions still dominate H and possibly He. **Figure 7** shows three events, sequentially from a single region rotating across the Sun, that show characteristic behavior of SEP2, then SEP3 events: temperature > 2 MK (like impulsive events) and enhanced high-$Z$ fit line that fails to include the proton intensities. **Figures 7D** and **7E** tend to show He enhanced as well and the high-$Z$ enhancements flatten with time, probably from preferential leakage at high $Z$, making temperature measurement difficult. The source rotates from W71 to W84 to W120 at the rate of ~13º day$^{-1}$. CME speeds for the three events are 830, 1199, and 2465 km s$^{-1}$ and the last two events are both GLEs. These "double dipping" SEP3 events are not uncommon (Reames 2022a) and often include GLEs. Of course, the GLE is determined by the protons, not the high-$Z$ ions, but the location or configuration of these events could be a factor.

While it is tempting to think that the single SEP2 event "feeds" the following SEP3 events in **Figure 7**, details of the abundances differ. The SEP2 event has unusually high Ne in **Figure 7C**, but the later SEP3 events in **Figures 7D** and **7E** do not. Presumably the seed-population for the SEP3 events is fed by many subsequent smaller impulsive events on 15 April that do not contribute energetic ions at 1 AU. SEP3 abundances always show smaller variations than SEP 2 events (e.g. figure 8 in Reames 2021b). Otherwise, searches for abundance features, like the Ne enhancement here, sometime implicate spectral fluctuations (Reames 2019). However, below 1 MeV amu$^{-1}$ Mason et al. (2016) found extreme spikes in S in 16 events in 16 years. The S may be a second-harmonic resonance at $A/Q = 3$ related to the $^3$He resonance at $A/Q = 1.5$ since S and $^3$He have similar spectra. In contrast, S ($Z = 16$) is actually suppressed in **Figure 3C**.

It is common that the high-$Z$ enhancement in SEP3 events is less than that of any preceding SEP2 events that may feed the impulsive pool, as in **Figure 7**. Shock acceleration probably reduces the enhancement from the seed impulsive ions just as modest SEP4 events have depressed the abundances of the ambient coronal seed ions in **Figure 4**.

**Figure 8** directly compares a SEP3 event with two SEP4 events. While the temperatures of the seed populations of the two event types differ, the most obvious difference occurs in the proton enhancements. The protons fit the power-law extrapolation from high $Z$ in **Figure 8F** but they are clearly enhanced in **Figure 8C** as they were in **Figure 7**. All three events in **Figure 8** show enhancements of high-$Z$ elements, at least initially, but it is the departure of the protons from the fit that suggests the presence of two seed populations in the SEP3 (and SEP2) events.

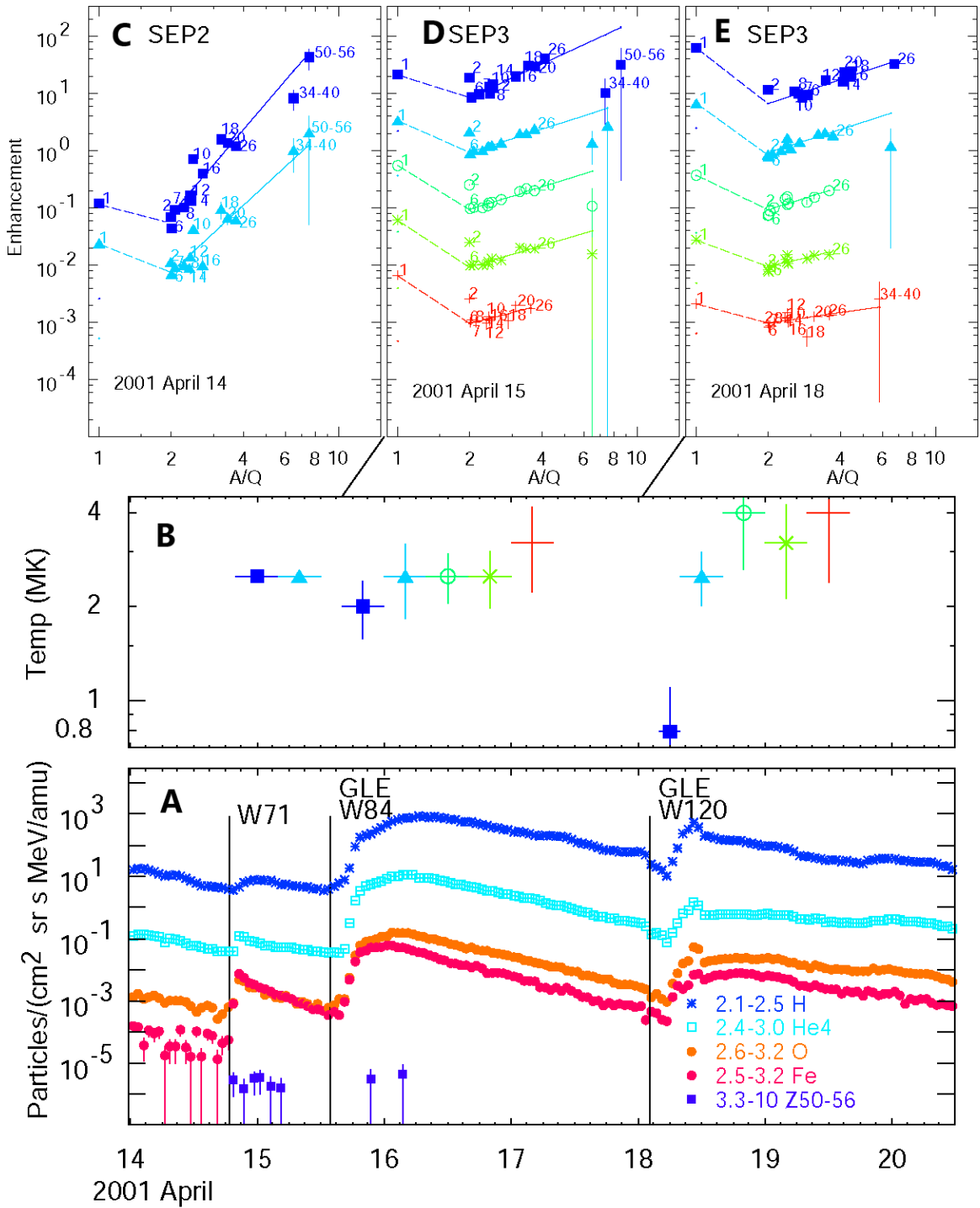

**FIGURE 7** Large GLEs as gradual SEP3 events. Panels **A** shows time evolution of the elements H, ⁴He, O, Fe, and elements with $50 \leq Z \leq 56$, along with listed energies (in MeV amu⁻¹) in two SEP3 events. Panel **B** shows derived source-plasma temperatures in a series of time intervals (**colors**) during the events. Panels **C, D,** and **E** show abundance enhancements vs. $A/Q$ for elements (normalized at O and shifted ×0.1 for each interval), with representative atomic numbers $Z$ shown, with best least-squares fits of elements with $Z \geq 6$, for each time interval (**color**) mapped to the events below. Two different seed populations dominate high and low $Z$ in SEP2 (**C**) and SEP3 (**D** and **E**) events.

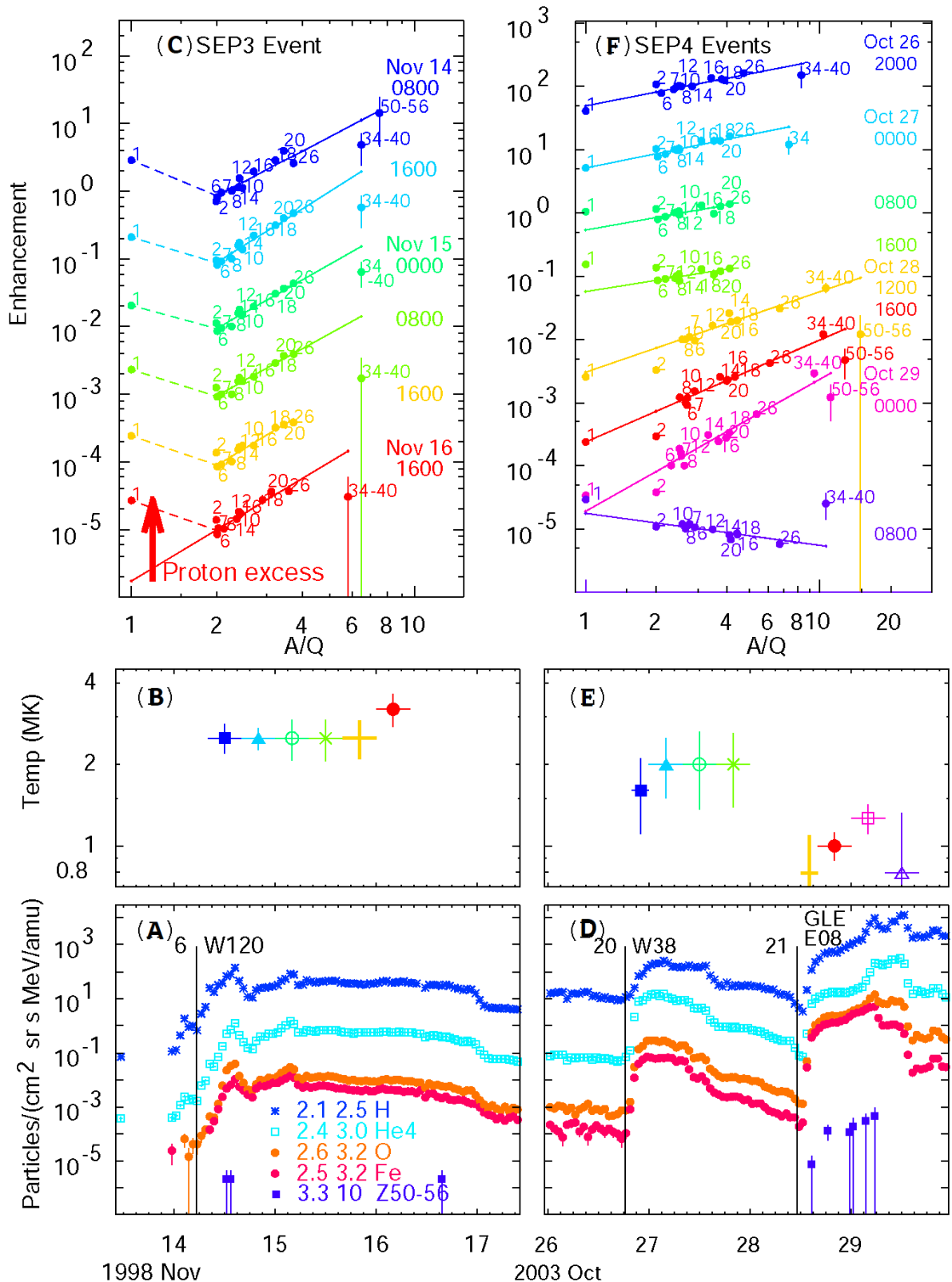

**FIGURE 8** Compares a large SEP3 (**left**) event with a pair of large SEP4 events (**right**); one is a GLEs. Panels **A** and **D** show time evolution of the elements H, ⁴He, O, Fe, and elements with $50 \leq Z \leq 56$, along with listed energies (in MeV amu⁻¹). Panels **B** and **E** show derived source-plasma temperatures in a series of intervals (**colors**) during each event. Panels **C** and **F** show abundance enhancements vs. $A/Q$ for elements (normalized at O and shifted ×0.1 for each interval) with $Z$ listed for each time interval (**color**) along with best least-squares fits of elements with $Z \geq 6$. Systematic proton excesses suggest the SEP3 event in **C** has 2 seed populations for the shock, while for SEP4 events in **F** all elements tend to fit a single power law.

It is important to realize that we cannot exclude the possibility of some impulsive suprathermal seeds in *any* gradual event, including SEP4 events. Mason et al. (1999) found a modest increase of $^3$He in a large gradual event. We cannot distinguish the presence of impulsive heavy elements unless they actually dominate in the event so we can see their characteristic *A/Q* pattern and higher temperature. If only ~10% of the Fe were from residual impulsive ions, we would never know it. It is also quite possible that very large events become SEP4 events because strong shocks sweep up enough ambient plasma to swamp any residual impulsive ions that are also available. It is also possible that SEP4 events occur because there are no residual SEP1 ions available for the shock to reaccelerate.

## 5 Conditions for SEP3 Events

GLEs are determined by proton intensities, and protons are accelerated from the ambient plasma in both SEP3 and SEP4 events, so GLE existence is independent of the dominant source of high-*Z* ions. In solar cycle 23, 6 of the 15 GLEs were SEP3 events and 9 were SEP4 events. Solar cycle 24 is much weaker with only 2 GLEs, and both were SEP4 events. During solar cycle 24, STEREO plus Earth provided three approximately equally spaced locations around the Sun. Cohen et al. (2017) observed H, He, O, and Fe for gradual SEP events seen by two or three spacecraft. Of a total of 41 events, 10 were measured on all three spacecraft. All of the 10 were SEP4 events and only one of the two-spacecraft events (4 August 2011) was a SEP3 event (see figure 10 in Reames 2020b).

Why are there so few SEP3 events in cycle 24? These events require a stream of residual impulsive SEP1 or SEP2 ions flowing out from an active region. These streams or persistent $^3$He-rich regions are often seen during solar maxima (Richardson et al. 1990; Desai et al. 2003; Wiedenbeck et al. 2008; Bučík et al. 2014, 2015; Chen et al. 2015; Reames 2022a). Then that same region emits a fast, wide CME-driven shock that accelerates these residual impulsive ions along with ambient ions that dominate the protons. In a weak solar cycle, (1) the number of fast CMEs is reduced and (2) the probability of $^3$He-rich streams is reduced, so perhaps the number of SEP3 events is reduced by the product of the two factors. Perhaps a very strong cycle would have mostly SEP3 events.

What are the conditions for producing a SEP3 event? Gopalswamy et al. (2022) ask an important question: "Can type III radio storms be a source of seed particles to shock acceleration?" These authors identify a large shock event that follows from the same active region as a storm of type III bursts. Should we expect a SEP3 event? **Figure 9** shows the time variation of SEP species during this period. He intensities show $^3$He/$^4$He ~ 1 during the type III storm, but intensities are too small to show measurable Fe and O. Intensities of $^3$He are not reliable during the large event because of background from $^4$He, but most likely $^3$He/$^4$He < 0.1. However, **Figures 9B** and **9C** show quite clearly that the large event is a SEP4 event and definitely not a SEP3 event. Source temperatures are quite low and protons fit the same population as the high-*Z* ions and the power-law fits actually tend to decline with *A/Q*. Impulsive suprathermal seed ions do not dominate this event.

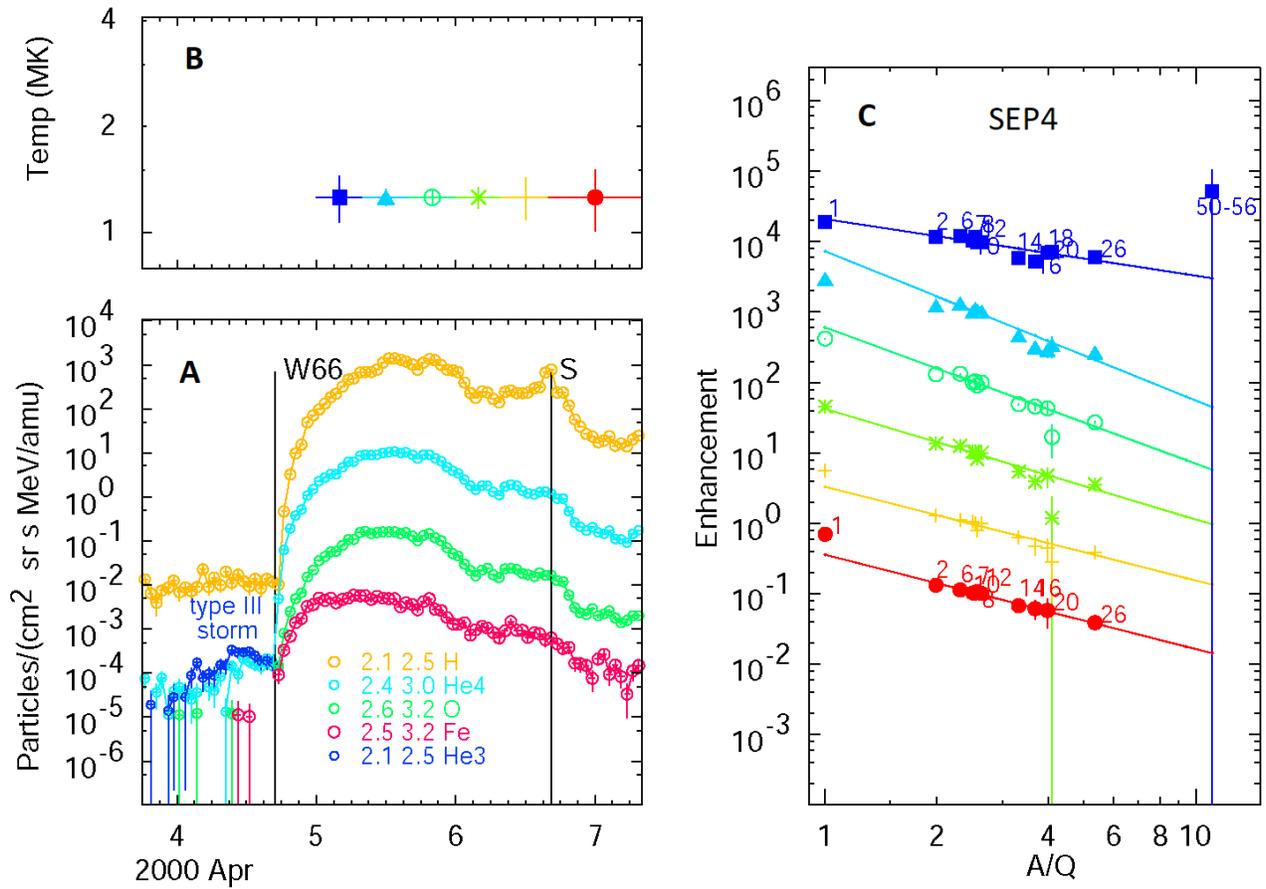

**FIGURE 9** Panel **A** shows intensities of SEP ions listed are shown during a type III radio storm and the following large SEP4 event from the same active region studied by Gopalswamy et al. (2022). **B** shows derived temperatures for selected intervals. Panel **C** shows abundance enhancements vs. $A/Q$ for elements with $Z$ listed for each time interval (**color and symbol**) along with best least-squares fits of elements with $Z \geq 6$ extended down to protons. Ions during the type III storm are $^3$He-rich but Fe/O is not measurable. $^3$He cannot be reliably measured during large event because of high $^4$He background. The large event is SEP4 and shows no evidence of impulsive seed ions from the type-III storm.

Unfortunately the intensities in **Figure 9** are insufficient for a firm conclusion since the type-III storm is too weak to show high-Z ions, but the large events is clearly not dominated by impulsive seeds, so it is unlikely that ions from the storm have contributed enough seed particles. There are other cases like this where large shock waves do not seem to reaccelerate available impulsive suprathermal ions; many large events are preceded by several type-III bursts. Are the intensities of impulsive seeds just too low? Yet there are also many cases, like **Figure 7**, where consecutive large shocks dip into a single persistent impulsive population (Reames 2022a). Worse, we have no cases where a single large event is SEP3 at one longitude and SEP4 at another; such cases might guide us to the location of the impulsive seeds, but there are too few in cycle 24. Perhaps, some shocks are driven in the direction of the impulsive seeds while others are driven away from them, but we do not know. The answer to the question posed by Gopalswamy et al. (2022) is not yet clear; impulsive seed ions are inadequate or there may be other required conditions. However, the association of $^3$He with type III emission remains strong.

In fact, the measurements of $^3$He/$^4$He during a type-III storm shown in **Figure 9A** address another important question: are all the electron events that produce type-III bursts $^3$He-rich? A value of $^3$He/$^4$He ≈ 1 persists during the entire type-III storm, suggesting that even these small events are $^3$He-rich. This may be the first reported measurement of $^3$He/$^4$He associated with the small events in a type-III storm. At the other extreme, $\gamma$-ray lines say that some of the largest flares are $^3$He-rich (Mandzhavidze et al., 1999; Murphy et al., 2016). Are $^3$He-rich events a persistent consequence of magnetic reconnection?

## 6    C/O and the Element Abundances of the Photosphere

After averaging over all the smooth power-law abundance variations, how could variation of a single abundance C/O stand out so dramatically in **Figure 1**? SEPs have C/O = 0.42±0.01, while the recent photospheric values are 0.550±0.76 (Caffau et al. 2011) and 0.589±0.063 (Asplund et al 2021). Earlier photospheric measurements of Anders and Grevesse (1989) of C/O=0.489 were lower and in better agreement. Since the FIP of C is lower than that of O, it seems extremely unlikely that C/O could be suppressed in transit to the corona. Mixing of various seed populations as described above can surely cause doubt in relative abundances of H and $^4$He, but the power-law behaviour of both transport and acceleration is most likely to preserve the relationship of closely-spaced C, N, and O. How could C/O in SEPs possibly be suppressed below that in the photosphere? We have previously suggested that the problem might rest with the coronal abundance of C (Reames 2021b).

It is well known that the increasing photospheric abundances of "heavy elements" like C, N, and O, have also caused problems for stellar models and helioseismology (e.g. Basu and Antia, 2008). These abundances determine the opacity of stellar material and the new (since 1990) lower abundances disagree with helioseismic constraints.

To correct the C/O problem and **Figure 1**, rather than decrease the photospheric C, as previously considered, suppose we raise photospheric O instead. While this is less convenient, since O is our reference, reducing the photospheric C/O to 0.42 amounts to reducing SEP O by 31% in Figure 1. At this point, C, O, Ne, and Ar, all line up. That is, their SEP and photospheric abundances are the same. If we assume that the SEP-derived photosphere has the observed SEP abundance ratios of all high-FIP elements (other than H and He), normalized to the abundance of C from Caffau et al. (2011), we find the values in Table 2. The higher abundance of O returns to the values in Anders and Grevesse (1989)

**Table 2** A SEP-based High-FIP Photosphere

|    | Z | FIP [eV] | SEPs | SEP Photo-sphere [dex] |
|----|---|----------|------|------------------------|
| C  | 6 | 11.3 | 420±10 | 8.50 |
| N  | 7 | 14.5 | 128±8 | 7.98 |
| O  | 8 | 13.6 | 1000±10 | 8.88 |
| Ne | 10 | 21.6 | 157±10 | 8.07 |
| Ar | 18 | 15.8 | 4.3±0.4 | 6.51 |

## 7    Discussion

Our analysis of source temperatures has assumed Maxwellian electron velocity distributions controlling the relationship between plasma temperatures and $Q$-values of ions (e.g. Mazzotta et al 1998). Recently, Lee et al. (2022, 2024) have tested this assumption using more-realistic kappa distributions, which include high-energy tails to the electron distribution, to fit average impulsive-event enhancements. These authors find that the derived source temperatures are not significantly affected and that $A/Q$-values differ at most by 10 – 20% in extreme cases. This is an important confirmation of the temperatures deduced from SEP abundances. These authors also note that the derived source temperatures agree with active-region temperatures, i.e. they have been unaffected by electron heating. Thus the SEPs must leave the acceleration region on a time scale shorter than their ionization time scales. However, at low densities these ionization times could be quite long.

A recent paper by Laming and Kuroda (2023) suggests that heavy ions in impulsive SEP events are enhanced as a part of the FIP process. Of course, the FIP process occurs in the dense chromosphere while ion acceleration must occur at much lower density in jets (Bučík 2020), decoupling enhancement and acceleration, but presumably the enhancement could affect a region that would later release a jet. However, this would suggest that both SEPs and the CME from a jet would have strong $A/Q$-dependent enhancements, which have never been seen for CMEs. It is also true than the high-FIP elements He, C, N, O, Ne, S, and Ar fit on the same pattern of enhancements as the low-FIP elements Mg, Si, Ca, and Fe (figure 2 in Reames 2023c; figure 8 in Reames et al. 2014a). Independently of the impulsive abundance enhancements, we have separate evidence that the acceleration occurs at magnetic reconnection sites in solar jets. Developing the power-law enhancement during this acceleration (Drake et al 2009) seems most likely, since only the SEPs would be affected.

However, concerning other mechanisms, we should note again that there is evidence that second-harmonic resonant processes related to enhancement of $^3$He may contribute to low-energy abundance enhancements especially of S in relatively rare, small impulsive events (Mason et al. 2016). These rare S-rich events have steep spectra that are only seen below ~1 MeV amu$^{-1}$ (Mason et al. 2016), probably where $A/Q \approx 3$ for S, which occurs at ~2 MK. These higher-order resonant enhancements associated with the $^3$He resonance at $A/Q \approx 1.5$. Resonant element enhancement at higher temperatures was studied by Roth and Temerin (1997). More recently, at even lower energies < 300 keV amu$^{-1}$, Mason et al. (2023) have found extreme enhancements in heavy elements, e.g. Fe/O, presumably because O has $A/Q \approx 2$, and these resonant waves have been absorbed by $^4$He. This is reminiscent of the hot-plasma study of "The He Valley" (Steinacker et al., 1997) where the wave absorption band of $^4$He can shape the abundances of other ions. Yet, all of the resonant modifications of the $Z > 6$ abundances were confined to steep spectra below ~1 MeV amu$^{-1}$, while the higher-energy abundances are dominated by power laws in $A/Q$.

There has been growing interest in the importance of streamers in SEP intensities and the production of GLEs. In streamers higher densities and lower Alfvén speeds produce higher Alfvénic Mach numbers (Liu et al. 2023); regions of higher $\theta_{Bn}$ (e.g. Kong et al. 2017, 2019) may also be a factor in shock acceleration. We have yet to explore any relationship between streamers and the streams of impulsive suprathermal ions above active regions that distinguish SEP3 and SEP4 events.

## 8    Summary

SEP element abundances relative to O are compared at the same velocity (i.e. MeV amu$^{-1}$) so that any dependence upon magnetic rigidity (often a power-law) appears as a dependence upon $A/Q$. Assuming a power law allows a best-fit determination of $Q$-values and hence source temperature.

Impulsive events are accelerated in magnetic-reconnection regions in solar jets and escape on open field lines from sources at ~2.5 MK. They produce $^3$He-rich events, heavy-ion enhancements and the electron beams that drive type III radio bursts. The smaller SEP1 events have no fast shocks available for additional acceleration. Steep power-law abundance enhancements vs. $A/Q$ (e.g. **Figure 3C**) include *all* elements, including H (Events 3 and 4) or, in some cases, begin above $^4$He (Event 5).

Gamma-ray line measurements tell us that solar flares involve the same $^3$He-rich acceleration mechanisms as solar jets and show significant nuclear fragmentation, but the high (>10 MK) temperatures and nuclear fragments of bright, hot flares are not found in SEPs, indicating the energetic ions in flares are efficiently trapped magnetically and are not observed to "leak out" into space. Thus, flares show no contribution to either impulsive or gradual SEPs.

In large gradual SEP4 events, fast, wide shock waves, driven by CMEs, are completely dominated by ions accelerated from ambient coronal seed material at 0.8-1.8 MK, beginning at 2 – 3 $R_S$. All elements, including H and $^4$He, again tend to fit on a single power law vs. $A/Q$. Slopes of the power-law abundance enhancements or depressions vs. $A/Q$ depend mainly upon scattering during transport. (1) Intensities in smaller SEP4 events produce minimal amplification of Afvén waves allowing high-rigidity ions to leak away preferentially so abundances decrease vs. $A/Q$ (**Figures 4C** and **4F**). (2) Intensities in larger SEP4 events produce some amplification of Afvén waves producing balanced trapping and leakage over the $A/Q$ range, so abundances are initially quite flat vs. $A/Q$ (**Figure 5C**). (3) Intensities of protons streaming out from very large SEP4 events (e.g. GLEs) produce significant amplification of Alfvén waves that scatter subsequent ions during transport out from the shock, but allow high-rigidity ions to preferentially to reach the observer ahead the shock, so abundances increase vs. $A/Q$ (**Figures 6C** and **6F**); depleted high-rigidity ions are seen as a depression vs. $A/Q$ downstream of the shock.

SEP events can include both shock-accelerated coronal seed ions and shock-reaccelerated impulsive residual seed ions. Impulsive SEP2 events are intended to involve both seeds and shock from a single jet while gradual SEP3 events describe a large shock traversing an active region with a collection of impulsive ions from multiple previous jets. Both classes are dominated by H (and possibly $^4$He) from the coronal seeds and by $Z \geq 6$ ions from the previously-enhanced impulsive seeds. **Figure 7C** shows the enhancement pattern of an impulsive SEP2 event, **Figures 7D, 7E** and **8C** show clear SEP3 events, and **Figure 3F** shows an ambiguous single-jet event (SEP2) that happens to occur in a pre-existing impulsively-seeded region (SEP3). SEP3 events tend to have smaller abundance fluctuations since they average over many individual jets.

SEP3 events are very rare in solar cycle 24, probably a combined effect of both fewer impulsive seeds and fewer large shocks to encounter them. Also, while type III storms are clearly found to produce $^3$He-rich events, subsequent strong shocks from that region do not necessarily find dominant high-$Z$ impulsive seed particles.

The SEP value of C/O = 0.42 conflicts with a much higher photospheric values up to 0.59 and lacks explanation. Could the photospheric value of O actually be 30 – 40% higher, as previously found, and as helioseismology independently suggests?

## 9    Conflict of Interest

The author declares that this research was conducted in the absence of any commercial or financial relationships that could be construed as a potential conflict of interest.

## 10    Author Contributions

All work on this article was performed by D.V. Reames.

## 11    Funding

No institutional funding was provided for this work.

## 12    References

Afanasiev, A., Battarbee, M., Vainio, R.: Self-consistent Monte Carlo simulations of proton acceleration in coronal shocks: Effect of anisotropic pitch-angle scattering of particles. Astron. Astrophys. **584**, 81 (2015). https://doi.org/10.1051/0004-6361/201526750

Afanasiev, A., Vainio, R., Trotta, D., Nyberg, S., Talebpour Sheshvan, N., Hietala, H., Dresing, N., Self-consistent modeling of the energetic storm particle event of November 10, 2012, Astron. Astrophys. 679 A111 (2023) https://doi.org/10.1051/0004-6361/202346220

Anders, E., Grevesse, N., Abundances of the elements: Meteoritic and solar. Geochim. Cosmochim. Acta **53**, 197 (1989). https://doi.org/10.1016/0016-7037(89)90286-X

Archontis, V., Hood, A.W., A numerical model of standard to blowout jets, Astrophys. J. Lett., **769** L21 (2013), doi: 10.1088/2041-8205/769/2/L21 [Erratum: Astrophys. J. Lett., 770, (2013), L41].

Asplund M., Amarsi A.M., Grevesse, N., The chemical make-up of the Sun: a 2020 vision. Astron. Astrophys. **653** A141 (2021). arXiv:2105.01661, doi: 10.1051/0004-6361/202140445

Basu, S., Antica, H.M., Helioseismology and solar abundances, Phys. Reports **457** 217 (2008) doi: 10.1016/j.physrep.2007.12.002

Bertsch, D.L., Fichtel, C.E., Reames, D.V., Relative abundance of iron-group nuclei in solar cosmic rays, Astrophys. J. Lett. **157**, L53 (1969) doi: 10.1086/180383

Breneman HH, Stone EC (1985) Solar coronal and photospheric abundances from solar energetic particle measurements. Astrophys J Lett 299:L57. https://doi.org/10.1086/184580

Bučík, R., ³He-rich solar energetic particles: solar sources, *Space Sci. Rev.* **216** 24 (2020) doi: 10.1007/s11214-020-00650-5

Bučík, R., Innes, D.E., Mall, U., Korth, A., Mason, G.M., Gómez-Herrero, R., Multi-spacecraft observations of recurrent ³He-rich solar energetic particles, Astrophys. J. **786**, 71 (2014) doi: 10.1088/0004-637X/786/1/71

Bučík, R., Innes, D.E., Chen, N.H., Mason, G.M., Gómez-Herrero, R., Wiedenbeck, M.E., Long-lived energetic particle source regions on the Sun, J. Phys. Conf. Ser. **642**, 012002 (2015) doi: 10.1088/1742-6596/642/1/012002

Bučík, R., Innes, D.E., Mason, G.M., Wiedenbeck, M.E., Gómez-Herrero, R., Nitta, N.V., ³He-rich solar energetic particles in helical jets on the Sun, Astrophys. J. **852** 76 (2018a) doi: 10.3847/1538-4357/aa9d8f

Bučík, R., Mulay, S.M., Mason, G.M., Nitta, N.V., Desai, M.I., Dayeh, M.A., Temperature in solar sources of ³He-rich solar energetic particles and relation to ion abundances, Astrophys. J. **908** 243 (2021) doi: 10.3847/1538-4357/abd62d

Bučík, R., Wiedenbeck, M.E., Mason, G.M., Gómez-Herrero, R., Nitta, N.V., Wang, L., ³He-rich solar energetic particles from sunspot jets, Astrophys. J. Lett. **869** L21 (2018b) doi: 10.3847/2041-8213/aaf37f

Caffau, E., Ludwig, H.-G., Steffen, M., Freytag, B., Bonofacio, P.,: Solar chemical abundances determined with a CO5BOLD 3D model atmosphere. Sol. Phys. 268, 255 (2011). https://doi.org/10.1007/s11207-010-9541-4

Chen N.H., Bučík R., Innes D.E., Mason G.M., Case studies of multi-day ³He-rich solar energetic particle periods, Astron. Astrophys. **580**, 16 (2015) doi: 10.1051/0004-6361/201525618

Cliver, E.W., Kahler, S.W., Reames, D.V., Coronal shocks and solar energetic proton events, Astrophys. J. **605**, 902 (2004) doi: 10.1086/382651

Cohen, C.M.S., Mason, G.M., Mewaldt, R.A., Characteristics of solar energetic ions as a function of longitude, Astrophys. J. **843** 132 (2017) doi: 10.3847/1538-4357/aa7513

Cook, W.R., Stone, E.C., Vogt, R.E., Elemental composition of solar energetic particles, Astrophys. J. **279**, 827 (1984) doi: 10.1086/161953

Desai, M., Giacalone, J., Large gradual solar energetic particle events, Living Reviews of Solar Physics (2016) doi: 10.1007/s41116-016-0002-5

Desai, M.I., Mason, G.M., Dwyer, J.R., Mazur, J.E., Gold, R.E., Krimigis, S.M., Smith, C.W., Skoug, R.M., Evidence for a suprathermal seed population of heavy ions accelerated by interplanetary shocks near 1 AU, Astrophys. J., 588, 1149 (2003) doi: 10.1086/374310

DiFabio, R., Guo, Z., Möbius, E., Klecker, B., Kucharek, H., Mason, G. M., et al. (2008). Energy-dependent charge states and their connection with ion abundances in impulsive solar energetic particle events. Astrophysical J. 687, 623–634. doi:10.1086/591833

Drake, J.F., Cassak, P.A., Shay, M.A., Swisdak, M., Quataert, E., A magnetic reconnection mechanism for ion acceleration and abundance enhancements in impulsive flares, Astrophys. J. Lett. **700**, L16 (2009) doi: 10.1088/0004-637X/700/1/L16

Fichtel, C.E., Guss, D.E., Heavy nuclei in solar cosmic rays, Phys. Rev. Lett. 6, 495 (1961) doi:  10.1103/PhysRevLett.6.495

Fisk, L.A., ³He-rich flares - a possible explanation, Astrophys. J. **224**, 1048 (1978) doi: 10.1086/156456

Forbush, S.E., Three unusual cosmic ray increases possibly due to charged particles from the Sun, Phys. Rev. **70**, 771 (1946) doi: 10.1103/PhysRev.70.771

Gopalswamy, N., Akiyama, S., Mäkelä, P., Yashiro, S., Hong Xie, H., Can type III radio storms be a source of seed particles to shock acceleration? 3rd URSI AT-AP-RASC, Gran Canaria, (2022) arXiv:2205.15852

Gopalswamy, N., Xie, H., Akiyama, S., Yashiro, S., Usoskin, I.G., Davila, J.M., The first ground level enhancement event of solar cycle 24: direct observation of shock formation and particle release heights, Astrophys J. Lett. **765** L30 (2013a) doi: 10.1088/2041-8205/765/2/L30

Gopalswamy, N., Xie, H., Mäkelä, P., Yashiro, S., Akiyama, S., Uddin, W., Srivastava, A.K., Joshi, N.C. Chandra, R., Manoharan, P.K., Mahalakshmi, K., Dwivedi, V.C., Jain, R., Awasthi, A.K., Nitta, N.V., Aschwanden, M.J., Choudhary, D.P., Height of shock formation in the solar corona inferred from observations of type II radio bursts and coronal mass ejections, Adv. Space Res. 51 1981 (2013b) doi: 10.1016/j.asr.2013.01.006

Gopalswamy, N., Xie, H., Yashiro, S., Akiyama, S., Mäkelä, P., Usoskin, I.G., Properties of Ground level enhancement events and the associated solar eruptions during solar cycle 23, Space Sci. Rev. **171**, 23 (2012) doi: 10.1007/s11214-012-9890-4

Gosling, J.T., The solar flare myth. J. Geophys. Res. **98**, 18937 (1993) doi: 10.1029/93JA01896

Gosling, J.T., Corrections to "The solar flare myth." J. Geophys. Res. **99**, 4259 (1994) doi: 10.1029/94JA00015

Kahler, S.W., Reames, D.V., Sheeley, N.R.,Jr., Coronal mass ejections associated with impulsive solar energetic particle events, Astrophys. J., **562**, 558 (2001) doi: 10.1086/323847

Kahler, S. W., Reames, D. V., Sheeley, N. R. Jr., Howard, R. A., Kooman, M. J., Michels, D. J., A comparison of solar $^3$He-rich events with type II bursts and coronal mass ejections, Astrophys, J. **290** 742 (1985) doi: 10.1086/163032

Kahler, S.W., Sheeley, N.R.,Jr., Howard, R.A., Koomen, M.J., Michels, D.J., McGuire R.E., von Rosenvinge, T.T., Reames, D.V., Associations between coronal mass ejections and solar energetic proton events, J. Geophys. Res. **89**, 9683 (1984) doi: 10.1029/JA089iA11p09683

Kahler, S.W., Tylka, A.J., Reames, D.V.: A comparison of elemental abundance ratios in SEP events in fast and slow solar wind regions. Astrophys. J. 701, 561 (2009). https://doi.org/10.1088/0004-637X/701/1/561

Kong, X., Guo, F. Giacalone, J., Li, H., Chen, Y., The acceleration of high-energy protons at coronal shocks: the effect of large-scale streamer-like magnetic field structures, Astrophys, J. **851** 38 (2017) doi: 10.3847/1538-4357/aa97d7

Kong, X., Guo, F. Chen, Y., Giacalone, J., The acceleration of energetic particles at coronal shocks and emergence of a double power-law feature in particle energy spectra, Astrophys. J. **883** 49 (2019) doi: 10.3847/1538-4357/ab3848

Kouloumvakos, A.,Mason, G.M., Ho, G.C., Allen, R.C., Wimmer-Schweingruber, R.F.,Rouillard, A.P., Rodriguez-Pacheco, J., Extended $^3$He-rich time periods observed by solar orbiter: magnetic connectivity and sources, Astrophys. J. **956** 123 (2023) https://doi.org/10.3847/1538-4357/acf44e

Kouloumvakos, A., Rouillard, A.P., Wu, Y., Vainio, R., Vourlidas, A., Plotnikov, I., Afanasiev, A., Önel, H, Connecting the properties of coronal shock waves with those of solar energetic particles, Astrophys. J. **876** 80 (2019) doi: 10.3847/1538-4357/ab15d7

Kozlovsky, B., Murphy, R.J., Ramaty, R.: Nuclear deexcitation gamma-ray lines from accelerated particle interactions. Astrophys. J. Suppl. 141, 523 (2002). https://doi.org/10.1086/340545

Laming, J.M.: The FIP and inverse FIP effects in solar and stellar coronae. Living Rev. Sol. Phys. 12, 2 (2015). https://doi.org/10.1007/lrsp-2015-2

Laming, J.M., Kuroda, N., Element abundances in impulsive solar energetic particle events, Astrophys. J. **951** 86 (2023) doi: 10.3847/1538-4357/acd69a.

Laming, J.M., Vourlidas, A., Korendyke, C., et al., Element abundances: a new diagnostic for the solar wind, Astrophys. J. **879** 124 (2019) doi: 10.3847/1538-4357/ab23f1  arXiv: 19005.09319

Lee, E.J., Archontis, V., Hood, A.W., Plasma jets and eruptions in solar coronal holes: a three-dimensional flux emergence experiment, Astrophys. J. Lett., **798** L10 (2015) doi: 10.1088/2041-8205/798/1/L10

Lee, J.Y., Kahler, S., Ko, Y.K., Raymond, J.C., A study of mass-to-charge ratio with various kappa values in impulsive SEP events, AGU Fall Meeting 2022, Chicago, IL, Bibcode: 2022AGUFMSH42A..04L

Lee, J.Y., Kahler, S., Raymond, J.C., Ko, Y.K, Solar energetic particle charge states and abundances with nonthermal electrons, Astrophys. J., in press (2024) arXiv:2401001604L

Lee, M.A.: Coupled hydromagnetic wave excitation and ion acceleration at interplanetary traveling shocks. J. Geophys. Res. 88, 6109 (1983) doi: 10.1029/JA088iA08p06109

Lee, M.A., Coupled hydromagnetic wave excitation and ion acceleration at an evolving coronal/interplanetary shock, Astrophys. J. Suppl., **158**, 38 (2005) doi: 10.1086/428753

Lee, M.A., Mewaldt, R.A., Giacalone, J., Shock acceleration of ions in the heliosphere, Space Sci. Rev. **173** 247 (2012) doi: 10.1007/s11214-012-9932-y

Lin, R.P.: The emission and propagation of 40 keV solar flare electrons. I: the relationship of 40 keV electron to energetic proton and relativistic electron emission by the sun. Sol. Phys. 12, 266 (1970). https://doi.org/10.1007/BF00227122

Liu, W., Kong, X., Guo, F. Zhao, L., Feng, S., Yu, F., Jiang, Z., Chen, Y., Giacalone, J., Effects of coronal magnetic field configuration on particle acceleration and release during the ground level enhancement events in solar cycle 24, Astrophys. J. **954** 203 (2023) doi: 10.3847/1538-4357/ace9d2  arXiv:2307.12191

Lodders, K., Palme, H., Gail, H.-P.: Abundances of the elements in the solar system, In: Trümper, J.E. (ed.) Landolt-Börnstein, New Series VI/4B, Chap. 4.4, p. 560. Springer, Berlin (2009). https://doi.org/10.1007/978-3-540-88055-4_34

Mandzhavidze, N., Ramaty, R., Kozlovsky, B., Determination of the abundances of subcoronal $^4$He and of solar flare-accelerated $^3$He and $^4$He from gamma-ray spectroscopy, Astrophys. J. **518**, 918 (1999) doi: 10.1086/307321

Mason, G.M.:$^3$He-rich solar energetic particle events. Space Sci. Rev. **130**, 231 (2007) doi: 10.1007/s11214-007-9156-8

Mason, G.M., Gloeckler, G., Hovestadt, D., Temporal variations of nucleonic abundances in solar flare energetic particle events. II - Evidence for large-scale shock acceleration, Astrophys. J. **280**, 902 (1984) doi: 10.1086/162066

Mason, G.M., Mazur, J.E., Dwyer, J.R., $^3$He enhancements in large solar energetic particle events, Astrophys. J. Lett. **525**, L133 (1999) doi: 10.1086/312349

Mason, G.M., Mazur, J.E., Dwyer, J.R., Jokippi, J.R., Gold, R.E., Krimigis, S.M., Abundances of heavy and ultraheavy ions in $^3$He-rich solar flares, Astrophys. J.  606, 555 (2004) doi: 10.1086/382864

Mason, G.M., Nitta, N.V., Wiedenbeck, M.E., Innes, D.E.: Evidence for a common acceleration mechanism for enrichments of 3He and heavy ions in impulsive SEP events. Astrophys. J. 823, 138 (2016). https://doi.org/10.3847/0004-637X/823/2/138

Mason, G.M., Reames, D.V., Klecker, B., Hovestadt, D., von Rosenvinge, T.T., The heavy-ion compositional signature in He-3-rich solar particle events, Astrophys. J. **303**, 849 (1986) doi: 10.1086/164133

Mason, G.M., Roth, I., Nitta, N.V., Bučík, R., Lario, D., Ho, G.C., Allen, R.C., Kouloumvakos, A,, Wimmer-Schweingruber, R.F., Rodriguez-Pacheco, J., Heavy-ion acceleration in $^3$He-rich solar energetic particle events: new insights from Solar Orbiter, Astrophys. J. **957** 112 (2023) https://doi.org/10.3847/1538-4357/acf31b

Mazzotta, P., Mazzitelli, G., Colafrancesco, S., Vittorio, N.: Ionization balance for optically thin plasmas: rate coefficients for all atoms and ions of the elements H to Ni. Astron. Astrophys Suppl. 133, 403 (1998). https://doi.org/10.1051/aas:1998330

McGuire, R.E., von Rosenvinge, T.T., McDonald, F.B., A survey of solar cosmic ray composition, *Proc. 16th Int. Cosmic Ray Conf., Tokyo* **5**, 61 (1979)

Melrose, D.B., *Plasma Astrophysics*, Vol. 1, (New York: Gordon and Breach) (1980)

Mewaldt, R.A., Cohen, C.M.S., Leske, R.A., Christian, E.R., Cummings, A.C., Stone, E.C., von Rosenvinge, T.T. and Wiedenbeck, M. E., Fractionation of solar energetic particles and solar wind according to first ionization potential, Advan. Space Res., **30**, 79 (2002) doi: 10.1016/S0273-1177(02)00263-6

Mewaldt, R.A., Looper, M.D., Cohen, C.M.S., Haggerty, D.K., Labrador, A.W., Leske, R.A., Mason, G.M., Mazur, J.E., von Rosenvinge, T.T., Energy spectra, composition, other properties of ground-level events during solar cycle 23, Space Sci. Rev. **171**, 97 (2012) doi: 10.1007/s11214-012-9884-2

Meyer, J.P., The baseline composition of solar energetic particles, Astrophys. J. Suppl. **57**, 151 (1985) doi: 10.1086/191000

Mogro-Campero, A., and Simpson, J. A. Enrichment of very heavy nuclei in the composition of solar accelerated particles. Astrophys. J. Lett. 171, L5 (1972) doi:10.1086/180856

Murphy, R.J., Ramaty, R., Kozlovsky, B., Reames, D.V., Solar abundances from gamma-ray spectroscopy: Comparisons with energetic particle, photospheric, and coronal abundances, Astrophys. J. **371**, 793 (1991) doi: 10.1086/169944

Murphy, R.J., Kozlovsky, B., Share, G.H.,: Evidence for enhanced ³He in flare-accelerated particles based on new calculations of the gamma-ray line spectrum, Astrophys.J. **833**, 166 (2016) doi: 10.3847/1538-4357/833/2/196

Ng, C.K., Reames, D.V.: Shock acceleration of solar energetic protons: the first 10 minutes, Astrophys. J. Lett. **686**, L123 (2008). https://doi.org/10.1086/592996

Ng, C.K., Reames, D.V., Tylka, A.J., Effect of proton-amplified waves on the evolution of solar energetic particle composition in gradual events, Geophys. Res. Lett. **26**, 2145 (1999) doi: 10.1029/1999GL900459

Ng, C.K., Reames, D.V., Tylka, A.J., Modeling shock-accelerated solar energetic particles coupled to interplanetary Alfvén waves, Astrophys. J. **591**, 461 (2003) doi: 10.1086/375293

Ng, C.K., Reames, D.V., Tylka, A.J., Solar energetic particles: shock acceleration and transport through self-amplified waves, AIP Conf. Proc. **1436**, 212 (2012) doi: 10.1063/1.4723610

Nitta, N.V., Reames, D.V., DeRosa, M.L., Yashiro, S., Gopalswamy, N., Solar sources of impulsive solar energetic particle events and their magnetic field connection to the earth, Astrophys. J. 650, 438 (2006) doi: 10.1086/507442

Pariat, E., Dalmasse, K., DeVore, C.R., Antiochos,S.K., Karpen, J.T., Model for straight and helical solar jets. I. Parametric studies of the magnetic field geometry, Astron. Astrophys. 573 A130 (2015) doi: 10.1051/0004-6361/201424209

Parker, E.N., The passage of energetic charged particles through interplanetary space, Planet. Space Sci. **13** 9 (1965) doi: 10.1016/0032-0633(65)90131-5

Post, D.E., Jensen, R.V., Tarter, C.B., Grasberger, W.H., Lokke, W.A.: Steady-state radiative cooling rates for low-density, high temperature plasmas. At. Data Nucl. Data Tables. 20, 397 (1977). https://doi.org/10.1016/0092-640X(77)90026-2

Raukunen, O., Vainio, R., Tylka, A.J., Dietrich, W.F., Jiggens, P., Heynderickx, D., Dierckxsens, M., Crosby, N., Ganse, U., Siipola, R.: Two solar proton fluence models based on ground level enhancement observations. J. Spa. Wea. Spa. Clim. 8, A04 (2018). https://doi.org/10.1051/swsc/2017031

Reames, D.V., Bimodal abundances in the energetic particles of solar and interplanetary origin, Astrophys. J. Lett. **330**, L71 (1988) doi: 10.1086/185207

Reames, D.V., Coronal Abundances determined from energetic particles, Adv. Space Res. **15** (7), 41 (1995a)

Reames, D.V., Solar energetic particles: A paradigm shift, Revs. Geophys. Suppl. **33**, 585 (1995b) doi: 10.1029/95RG00188

Reames, D.V., Particle acceleration at the Sun and in the heliosphere, Space Sci. Rev. **90**, 413 (1999) doi: 10.1023/A:1005105831781

Reames, D.V., Abundances of trans-iron elements in solar energetic particle events, Astrophys. J. Lett. **540**, L111 (2000) doi: 10.1086/312886

Reames, D. V., Solar release times of energetic particles in ground-level events, Astrophys. J. **693**, 812 (2009a) doi: 10.1088/0004-637X/693/1/812

Reames, D. V., Solar energetic-particle release times in historic ground-level events, Astrophys. J. **706**, 844 (2009b) doi; 10.1088/0004-637X/706/1/844

Reames, D.V., The two sources of solar energetic particles, Space Sci. Rev. **175**, 53 (2013) doi: 10.1007/s11214-013-9958-9

Reames, D.V., Element abundances in solar energetic particles and the solar corona, Solar Phys., **289**, 977 (2014) doi: 10.1007/s11207-013-0350-4

Reames, D.V.: Temperature of the source plasma in gradual solar energetic particle events. Sol. Phys. 291, 911 (2016). https://doi.org/10.1007/s11207-016-0854-9

Reames, D.V., "The FIP effect" and the origins of solar energetic particles and of the solar wind, Solar Phys. **293** 47 (2018a) doi: https://doi.org/10.1007/s11207-018-1267-8 (arXiv 1801.05840 )

Reames, D.V., Abundances, ionization states, temperatures, and FIP in solar energetic particles, Space Sci. Rev. **214** 61 (2018b) doi: 10.1007/s11214-018-0495-4

Reames, D.V., Helium suppression in impulsive solar energetic-particle events. Sol. Phys. **294**, 32 (2019). https://doi.org/10.1007/s11207-019-1422-x. (arXiv: 1812.01635)

Reames, D.V., Four distinct pathways to the element abundances in solar energetic particles, Space Sci. Rev.**216** 20 (2020a) doi: 10.1007/s11214-020-0643-5

Reames, D.V.: Distinguishing the rigidity dependences of acceleration and transport in solar energetic particles. Sol. Phys. **295**, 113 (2020b). https://doi.org/10.1007/s11207-020-01680-6. arXiv 2006.11338

Reames D.V., *Solar Energetic Particles*, (Second Edition) *Lec. Notes Phys. 978* Springer Nature, Cham, Switzerland, open access (2021a), doi: https://doi.org/ 10.1007/978-3-030-66402-2

Reames, D.V., Sixty years of element abundance measurements in solar energetic particles, Space Sci. Rev. **217** 72 (2021b) doi: 10.1007/s11214-021-00845-4

Reames, D.V., Fifty Years of ³He-rich Events, Front. Astron. Space Sci. **8**, 164. (2021c) doi:10.3389/fspas.2021.760261

Reames, D.V., The correlation between energy spectra and element abundances in solar energetic particles. Solar Phys. **296**, 24 (2021d) https://doi.org/10.1007/s11207-021-01762-z

Reames D.V., Energy spectra vs. element abundances in solar energetic particles and the roles of magnetic reconnection and shock acceleration. Sol Phys 297:32 (2022a). https://doi.org/10.1007/s11207-022-01961-2

# Element Abundances in Solar Energetic Particles

Reames, D.V., Solar energetic particles: spatial extent and implications of the H and He abundances, Space Sci. Rev. **218** 48 (2022b) doi; [10.1007/s11214-022-00917-z](10.1007/s11214-022-00917-z)

Reames, D.V., How do shock waves define the space-time structure of gradual solar energetic-particle events? Space Sci. Rev. **219** 14 (2023a) doi: [10.1007/s11214-023-00959-x](10.1007/s11214-023-00959-x)

Reames, D.V., Review and outlook of solar-energetic-particle measurements on multispacecraft missions, Front. Astron. Space Sci. **10** (2023b) [https://doi.org/10.3389/fspas.2023.1254266](https://doi.org/10.3389/fspas.2023.1254266)

Reames, D.V., Element abundances in impulsive solar energetic-particle events, Universe 9 466 (2023c) doi: [10.3390/universe9110466](10.3390/universe9110466) arXiv: [2309.09327](2309.09327)

Reames, D.V. Ng, C.K., Streaming-limited intensities of solar energetic particles, Astrophys. J. **504**, 1002 (1998) doi: [10.1086/306124](10.1086/306124)

Reames, D.V., Ng, C.K., Heavy-element abundances in solar particle events, Astrophys. J. **610**, 510 (2004) doi: 10.1088/0004-637X/723/2/1286

Reames, D.V., Ng, C.K., Streaming-limited intensities of solar energetic particles on the intensity plateau, Astrophys. J. **723** 1286 (2010) doi: [10.1088/0004-637X/723/2/1286](10.1088/0004-637X/723/2/1286)

Reames, D.V., Ng, C.K., The streaming limit of solar energetic-particle intensities, Living with a Star Workshop on Extreme Space Weather Events, Boulder, Co, June 9-11 (2014) arXiv [1412.2279](1412.2279)

Reames, D.V., Richardson, I.G., Barbier, L.M.: On the differences in element abundances of energetic ions from corotating events and from large solar events. Astrophys. J. Lett. **382**, L43 (1991). [https://doi.org/10.1086/186209](https://doi.org/10.1086/186209)

Reames, D.V., Stone, R.G., The identification of solar ³He-rich events and the study of particle acceleration at the sun, Astrophys. J., **308**, 902 (1986) doi: 10.1086/164560

Reames, D.V., Cliver, E.W., Kahler, S.W., Abundance enhancements in impulsive solar energetic-particle events with associated coronal mass ejections, Solar Phys. **289**, 3817, (2014a) doi: 10.1007/s11207-014-0547-1

Reames, D.V., Cliver, E.W., Kahler, S.W., Variations in abundance enhancements in impulsive solar energetic-particle events and related CMEs and flares, Solar Phys. **289**, 4675 (2014b) doi: 10.1007/s11207-014-0589-4

Reames, D.V., Meyer, J.P., von Rosenvinge, T.T., Energetic-particle abundances in impulsive solar flare events, Astrophys. J. Suppl. **90**, 649 (1994) doi: 10.1086/191887

Reames, D.V., von Rosenvinge, T.T., Lin, R.P., Solar ³He-rich events and nonrelativistic electron events - A new association, Astrophys. J. **292**, 716 (1985) doi: [10.1086/163203](10.1086/163203)

Richardson, I.G., Reames, D.V., Wenzel, K.P., Rodriguez-Pacheco, J., Quiet-time properties of < 10 MeV/n interplanetary ions during solar maximum and minimum, Astrophys. J. Lett. **363** L9 (1990) doi: [10.1086/185853](10.1086/185853)

Roth, I., Temerin, M.: Enrichment of ³He and heavy ions in impulsive solar flares. Astrophys. J. 477, 940 (1997). [https://doi.org/10.1086/303731](https://doi.org/10.1086/303731)

Serlemitsos, A.T., Balasubrahmanyan, V.K., Solar particle events with anomalously large relative abundance of ³He, Astrophys. J. **198**, 195, (1975) doi: [10.1086/153592](10.1086/153592)

Shimojo, M., Shibata, K., Physical parameters of solar X-ray jets, Astrophys. J. **542**, 1100 (2000) doi: [10.1086/317024](10.1086/317024)

Steinacker, J., Meyer, J.-P., Steinacker, A., Reames, D.V.: The helium valley: comparison of impulsive solar flare ion abundances and gyroresonant acceleration with oblique turbulence in a hot multi-ion plasma. Astrophys. J. 476, 403 (1997). https://doi.org/10.1086/303589

Stix, T.H., *Waves in Plasmas* (New York: AIP) (1992)

Tan, L.C., Reames, D.V., Ng, C.K., Shao, X., Wang, L.: What causes scatter-free transport of non-relativistic solar electrons? Astrophys. J. **728**, 133 (2011). [https://doi.org/10.1088/0004-637X/728/2/133](https://doi.org/10.1088/0004-637X/728/2/133)

Teegarden, B.J., von Rosenvinge, T.T., McDonald, F.B., Satellite measurements of the charge composition of solar cosmic rays in the 6 ≤ Z ≤ 26 interval. Astrophys. J. 180, 571 (1973). [https://doi.org/10.1086/151985](https://doi.org/10.1086/151985)

Temerin, M., Roth, I., The production of ³He and heavy ion enrichment in ³He-rich flares by electromagnetic hydrogen cyclotron waves, Astrophys. J. Lett. **391**, L105 (1992) doi: 10.1086/186408

Tylka, A.J., Dietrich, W.F.: A new and comprehensive analysis of proton spectra in ground-level enhanced (GLE) solar particle events. In: Proceedings of 31st International Cosmic Ray Conference Lodz. [http://icrc2009.uni.lodz.pl/proc/pdf/icrc0273.pdf](http://icrc2009.uni.lodz.pl/proc/pdf/icrc0273.pdf) (2009)

Tylka, A.J., Cohen, C. M. S., Dietrich, W. F., Krucker, S., McGuire, R. E., Mewaldt, R. A., Ng. C. K., Reames, D. V., Share, G. H., in *Proc. 28th Int. Cosmic Ray Conf.* (Tsukuba) **6** 3305 (2003)

Tylka, A.J., Cohen, C.M.S., Dietrich, W.F., Lee, M.A., Maclennan, C.G., Mewaldt, R.A., Ng, C.K., Reames, D.V., Shock geometry, seed populations, and the origin of variable elemental composition at high energies in large gradual solar particle events, Astrophys. J. **625**, 474 (2005) doi: 10.1086/429384

Tylka, A.J., Cohen, C.M.S., Dietrich, W.F., Maclennan, C.G., McGuire, R.E., Ng, C.K., Reames, D.V., Evidence for remnant flare suprathermals in the source population of solar energetic particles in the 2000 bastille day event, Astrophys. J. Lett. **558** L59 (2001) doi: [10.1086/323344](10.1086/323344)

Tylka, A.J., Lee, M.A., Spectral and compositional characteristics of gradual and impulsive solar energetic particle events, Astrophys. J. **646**, 1319 (2006) doi: 10.1086/505106

von Rosenvinge, T.T., Barbier, L.M., Karsch, J., Liberman, R., Madden, M.P., Nolan, T., Reames, D.V., Ryan, L., Singh, S., Trexel, H.: 1995, The energetic particles: acceleration, composition, and transport (EPACT) investigation on the wind spacecraft. Space Sci. Rev. 71, 152. doi: 10.1007/BF00751329

Wang, Y.-M., Pick, M., Mason, G.M., Coronal holes, jets, and the origin of ³He-rich particle events, Astrophys. J. **639**, 495 (2006) doi: [10.1086/499355](10.1086/499355)

Webber WR (1975) Solar and galactic cosmic ray abundances – a comparison and some comments. In: Proc. 14th Int. Cos. Ray Conf., Munixxh, vol 5, p 1597

Wiedenbeck, M. E., Cohen, C. M. S., Cummings, A. C., de Nolfo, G. A., Leske, R. A., Mewaldt, R. A., et al. (2008). Persistent Energetic ³He in the Inner Heliosphere. Proc. 30th Int. Cosmic Ray Conf. (Mérida) 1, 91.

Wild, J.P., Smerd, S.F., Weiss, A.A., Solar Bursts, Annu. Rev. Astron. Astrophys., **1**, 291 (1963) doi: [10.1146/annurev.aa.01.090163.001451](10.1146/annurev.aa.01.090163.001451)

Zank, G.P., Li, G., Verkhoglyadova, O., Particle Acceleration at Interplanetary Shocks, Space Sci. Rev. **130**, 255 (2007) doi: [10.1007/s11214-007-9214-2](10.1007/s11214-007-9214-2)

Zank, G.P., Rice, W.K.M., Wu, C.C., Particle acceleration and coronal mass ejection driven shocks: A theoretical model, J. Geophys. Res., 105, 25079 (2000) doi: 10.1029/1999JA000455